\documentclass[]{pasj02} 
\usepackage[switch,mathlines]{lineno} 
\usepackage{rotating}
\usepackage{multirow}
\usepackage{booktabs}
\usepackage{threeparttable}
\usepackage{xcolor}
\usepackage{natbib}
\makeatletter
\let\pasj@ssect\@ssect
\makeatother
\usepackage[
  bookmarks=false,
  colorlinks=true,
  citecolor=blue
]{hyperref}
\makeatletter
\let\@ssect\pasj@ssect
\makeatother

\newcommand*{\firstRev}[2][black]{\textcolor{#1}{#2}}
\definecolor{minorRev}{RGB}{0,0,0}

\jyear{2024}
\Received{}
\Accepted{}

\begin{document} 

\title{ 
XRISM view of a featureless cluster Abell 1060: \\
Revealing the advection of a diffuse radio source by ICM motions
}

\author{
Kohei \textsc{kurahara},\altaffilmark{1}\altemailmark\orcid{0000-0003-2955-1239} \email{kurahara.kohei.i7@f.mail.nagoya-u.ac.jp} 
Kazuhiro \textsc{nakazawa},\altaffilmark{1,2}\orcid{0000-0003-2930-350X}
Yuki \textsc{omiya},\altaffilmark{3,2}\orcid{0009-0009-9196-4174}
Daisuke \textsc{ito},\altaffilmark{2}\orcid{0009-0000-4742-5098}
Kosuke \textsc{nishiwaki},\altaffilmark{4}\orcid{0000-0003-2370-0475}
Andrea \textsc{botteon},\altaffilmark{4}\orcid{0000-0002-9325-1567}
Takuya \textsc{akahori},\altaffilmark{5,6}\orcid{0000-0001-9399-5331}
Noriko Y. \textsc{yamasaki},\altaffilmark{3}\orcid{0000-0003-4885-5537}
Liyi \textsc{gu},\altaffilmark{7}\orcid{0000-0001-9911-7038}
Kosuke~\textsc{sato},\altaffilmark{8}\orcid{0000-0001-5774-1633}
Takayuki \textsc{tamura},\altaffilmark{3}\orcid{0000-0003-2566-7541}
}
\altaffiltext{1}{Kobayashi-Maskawa Institute for the Origin of Particles and the Universe (KMI), Furo-cho, Chikusa-ku, Nagoya, Aichi 464-8601, Japan}
\altaffiltext{2}{Department of Physics, Nagoya University, Furo-cho, Chikusa-ku, Nagoya, Aichi 464-8601, Japan}
\altaffiltext{3}{Institute of Space and Astronautical Science (ISAS), Japan Aerospace Exploration Agency (JAXA), Kanagawa 252-5210, Japan}
\altaffiltext{4}{Istituto di Radioastronomia, INAF, Via Gobetti 101, 40121 Bologna, Italy}
\altaffiltext{5}{Mizusawa VLBI Observatory, National Astronomical Observatory of Japan, 2-21-1 Osawa, Mitaka, Tokyo 181-8588, Japan}
\altaffiltext{6}{Department of Astronomical Science, SOKENDAI, Shonan Village, Hayama-machi, Miura-gun, Kanagawa 240-0193, Japan}
\altaffiltext{7}{SRON Space Research Organisation Netherlands, Niels Bohrweg 4, 2333 CA Leiden, The Netherlands}
\altaffiltext{8}{Department of Astrophysics and Atmospheric Sciences, Kyoto Sangyo University, Motoyama, Kamigamo, Kita-ku, Kyoto, Kyoto 603-8555, Japan}



\maketitle

\begin{abstract}
Abell~1060 is a nearby, dynamically relaxed galaxy cluster hosting the extended radio structure Flying Fox (FF) near its center, providing a unique laboratory for studying intracluster turbulence and nonthermal phenomena. We used high-resolution X-ray spectroscopy with the XRISM/Resolve instrument to measure the velocity structure and velocity dispersion of the intracluster medium (ICM) in central Abell~1060. Spatially resolved analysis revealed a systematic velocity gradient of approximately $100~{\rm km~s^{-1}}$ along FF and a localized enhancement in velocity dispersion near the brightest cluster galaxy NGC~3311, suggesting turbulence generated by galaxy--ICM interactions. Together with the relatively small extent of FF compared with radio halos, these results favor fossil plasma supplied by past activity of the brightest cluster galaxy and transported by cluster-scale ICM motions. We also examined whether the observed turbulence could generate or maintain the nonthermal components of FF. For the measured velocity dispersion of $96.7^{+23.9}_{-24.6}~{\rm km~s^{-1}}$ and an assumed turbulence scale of $50$~kpc, the turbulent energy flux was estimated to be approximately $3.49 \times 10^{-28}~{\rm erg~cm^{-3}~s^{-1}}$. This scale is approximately half the $\beta$-model core radius and is comparable to the XRISM/Resolve field of view. Although this energy flux may weakly reenergize pre-existing relativistic electrons, the acceleration timescale of approximately $2.5$~Gyr exceeds the synchrotron and inverse Compton cooling timescale of approximately $210$~Myr. The present turbulence is therefore insufficient to accelerate electrons to high energies. In situ turbulent reacceleration is unlikely to be the dominant mechanism generating FF, although it may help maintain fossil electrons supplied by past AGN activity. These results provide observational evidence directly linking ICM dynamics to the origin of its nonthermal components and support a time-evolutionary picture of particle acceleration in galaxy clusters.
\end{abstract}


\section{Introduction}\label{sec:1}

Galaxy clusters are the largest gravitationally bound systems in the universe, and they are predominantly filled with the intracluster medium (ICM), a hot plasma with temperatures reaching several keV. While the ICM is primarily observed through its thermal X-ray emission, it is also known to host nonthermal components arising from synchrotron radiation and inverse Compton scattering \citep{l.sarazin1986xray}. In particular, diffuse radio emission provides direct observational evidence for the presence of relativistic electrons and magnetic fields, offering important clues to the history of energy transport and the role of turbulence in shaping the ICM. While various structures, such as radio halos and relics, have been extensively reported \citep{2019SSRv..215...16V}, recent advances in low-frequency radio instrumentation have revealed a wider diversity of morphologies \citep[e.g.,][]{2022Natur.609..911C, 2019Sci...364..981G}. Since their origins remain largely unclear, the physical processes that generate, maintain, and amplify these nonthermal components---especially particle acceleration and magnetic field amplification---remain under active debate.

One of the widely discussed models is turbulent reacceleration within galaxy clusters \citep{2001MNRAS.320..365B, 2015ApJ...815..116F, 2024ApJ...961...15N}. In this scenario, turbulence driven by cluster mergers and ICM motions is thought to reaccelerate pre-existing seed electrons, thereby producing the observed synchrotron emission. However, direct observational constraints remain limited on the strength and spatial scale of the turbulence, as well as on whether the acceleration timescale is consistent with the radiative cooling timescale in the ICM environment.

One of the fundamental reasons for these issues is that it has been difficult to directly measure turbulence and gas motions in the ICM. Previous X-ray observations have not been capable of measuring velocity structures and velocity dispersions with high precision, and the presence of turbulence has therefore been inferred mainly from indirect indicators. In this context, high-resolution X-ray spectroscopy with the X-Ray Imaging and Spectroscopy Mission \citep[XRISM;][]{2025PASJ...77S...1T} Resolve instrument provides a new means to directly measure the velocity dispersion from line broadening and bulk motions from shifts in line centroids, enabling constraints on the kinematic properties of the ICM with unprecedented precision \citep[e.g.,][]{2025Natur.638..365X, 2026PASJ...78..719O}.

In this study, we use XRISM/Resolve observations of the central region of Abell~1060 to measure the velocity structure and velocity dispersion of the ICM with high precision. Abell~1060 is known as a relatively relaxed galaxy cluster \citep{2000ApJ...535..602T, 2001PASJ...53..401M, 2001PASJ...53..421F, 2002ApJ...578..833Y, 2006PASJ...58..695H, 2007PASJ...59..299S, 2026A&A...711A.226S}; however, recent radio observations have revealed a distinctive extended radio structure, "Flying Fox" (hereafter FF), in its central region \citep{2024PASJ...76L...8K}. To understand the origin of this nonthermal emission structure, we investigate the turbulent properties of the ICM based on velocity measurements and assess their implications for particle acceleration and magnetic field amplification, based on \cite{2022SciA....8.7623B}. We also examine the global gas motions and local velocity structures through spatially resolved spectral analysis, and explore their connection to the spatial distribution of the nonthermal emission.

The structure of this paper is as follows. In \firstRev[minorRev]{section}~2, we describe the observations and data analysis methods. In \firstRev[minorRev]{section}~3, we present the results of the spectral analysis with XRISM/Resolve. In \firstRev[minorRev]{section}~4, we discuss the dynamical state of the ICM and the origin of the observed radio structure, followed by quantitative tests of particle reacceleration and magnetic field amplification due to turbulence. Finally, we summarize our conclusions in \firstRev[minorRev]{section}~5.

We adopt cosmological parameters of $H_0 = 70~{\rm km~s^{-1}~Mpc^{-1}}$, $\Omega_M = 0.3$, and $\Omega_\Lambda = 0.7$ throughout this paper.

\begin{figure*}[tp]
\begin{center}
\includegraphics[width=0.49\linewidth]{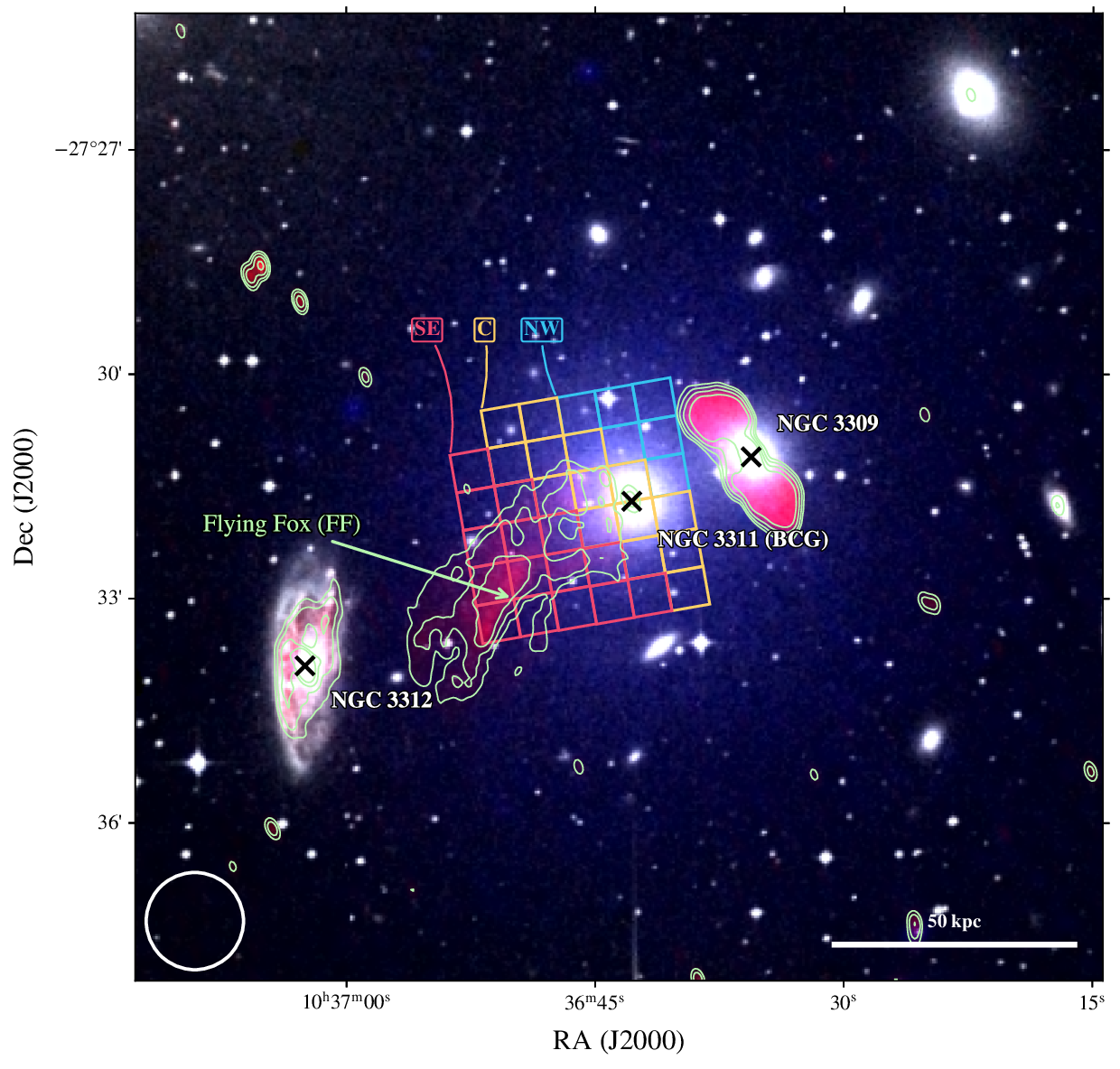}
\includegraphics[width=0.49\linewidth]{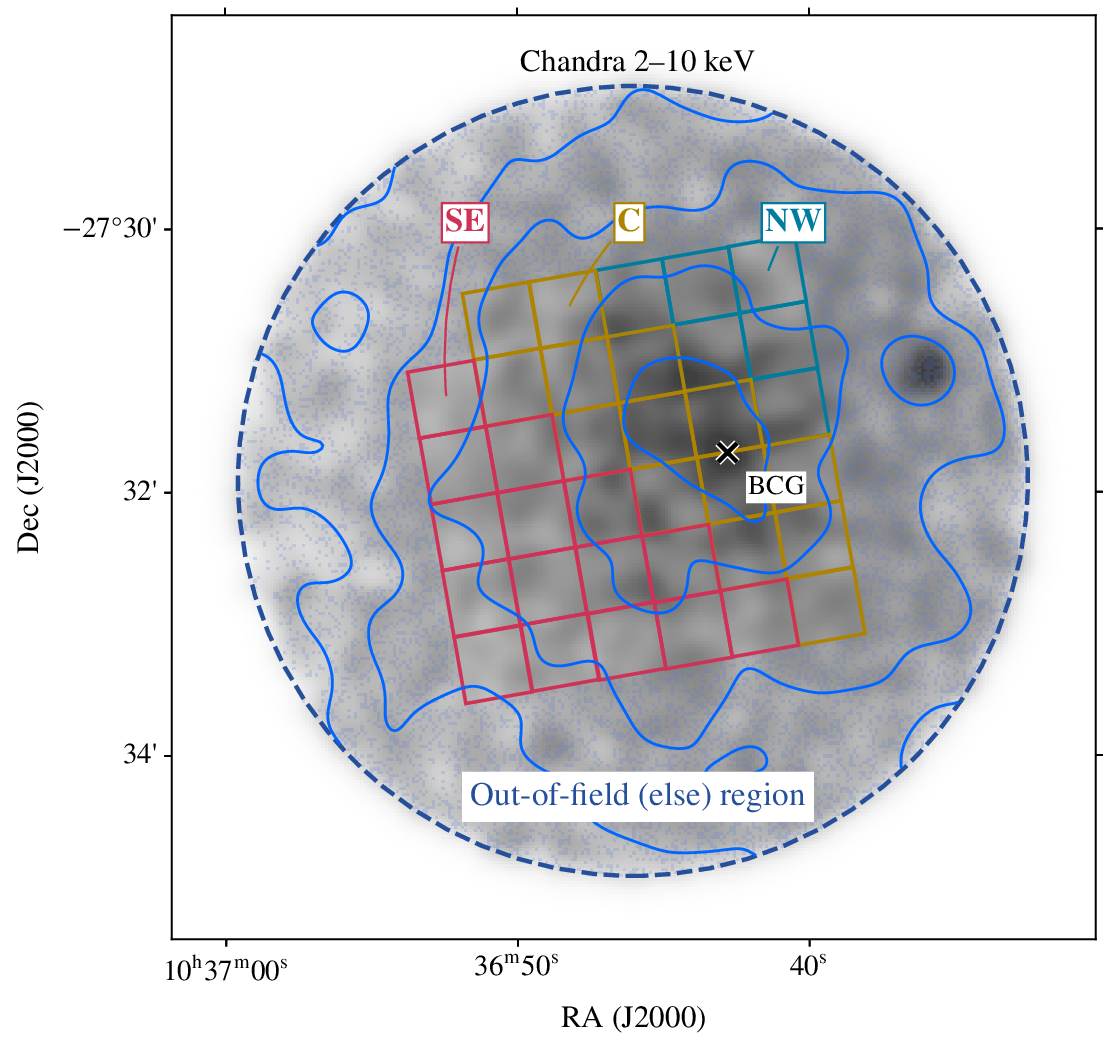}
\end{center}
\caption{\firstRev[minorRev]{(Left) Composite image of Abell~1060. White, blue, and red represent optical, X-ray, and radio emission, respectively. The colored pixel outlines indicate the three regions used for spatially resolved spectroscopy: SE (pink), C (yellow), and NW (cyan). Pale green contours trace the radio brightness at 0.5, 1, 2, and 4 mJy beam$^{-1}$; the arrow identifies FF. Black crosses mark NGC~3309, NGC~3311 (the BCG), and NGC~3312. The lower-left circle indicates the 1.3 arcmin half-power diameter of the Resolve PSF; the scale bar corresponds to 50 kpc. (Right) Chandra 2.0--10.0 keV brightness templates used in the SSM analysis, smoothed with a Gaussian of $\sigma=5.9\arcsec$ and displayed with an adjusted intensity scale. Bright blue contours trace the parent X-ray image smoothed with $\sigma=11.8\arcsec$, at 35, 45, 55, 70, and 85\% of its peak within the dashed circle. The same three in-field regions are outlined. Blue-tinted pixels outside these regions show the actual out-of-field template; the dashed circle shows its approximately 3 arcmin outer radius. The cross marks the BCG.}\\
\firstRev[minorRev]{Alt text: Left: optical, X-ray, and radio composite with labeled galaxies, three Resolve extraction regions, and radio contours identifying Flying Fox. Right: Chandra image with bright blue X-ray brightness contours, the same pixel regions, and the exterior emission template within a dashed circle.}
} 
\label{fig:rgb_combined_whiteDSS}
\end{figure*}

\section{Observation and data reduction}\label{sec:2}

We use XRISM/Resolve observations of the central region of Abell~1060 (obsID: 201043010, PI: Kazuhiro Nakazawa). The pointing center corresponds to J2000 coordinates of RA = 159.1922 deg and Dec = $-27.5305$ deg. The observation started on 2025 May 8 at 07:08:04 (UT). The data analyzed in this work were processed with pipeline version 03.01.014.010 and generated on 2025 May 23.

\firstRev[minorRev]{For comparison, the characteristic cluster radius is $R_{500}=45.38\arcmin$ \citep{2026A&A...711A.226S}, or approximately 710 kpc for a scale of $0.26~{\rm kpc\,arcsec^{-1}}$. The $3\arcmin\times3\arcmin$ Resolve FoV covers approximately $47\times47$ kpc: its half-width is only $0.033R_{500}$. Relative to a representative core radius of approximately 100 kpc, consistent with the $\beta$-model value $r_c=7.3\arcmin\pm1.8\arcmin$ measured by \citet{2006PASJ...58..695H} (approximately $114\pm28$ kpc at our adopted angular scale), the observation thus samples the inner quarter of the core radius along each side, rather than the whole cluster core.}

The basic data reduction procedures in this study follow the XRISM Data Reduction Guide (``ABC Guide'').
\footnote{%
  \href{https://heasarc.gsfc.nasa.gov/docs/xrism/analysis/abc_guide/xrism_abc.html}{%
    \nolinkurl{https://heasarc.gsfc.nasa.gov/docs/xrism/analysis/abc_guide/xrism_abc.html}%
  }
}
We further filtered the event files by selecting events with pulse invariant (PI) $\geq 600$ and nominal pulse rise-time properties, and extracted only the high-resolution primary (Hp) events. The analysis was performed using HEASoft version 6.35, and the calibration files for XRISM/Resolve were taken from CALDB version 20250915.

To estimate the non-X-ray background (NXB), \firstRev[minorRev]{we followed the Resolve NXB recipe}.\footnote{%
  \href{https://heasarc.gsfc.nasa.gov/docs/xrism/analysis/nxb/resolve_nxb_db.html}{%
    \nolinkurl{https://heasarc.gsfc.nasa.gov/docs/xrism/analysis/nxb/resolve_nxb_db.html}%
  }
}
\firstRev[minorRev]{Following this procedure,} we generated good time intervals (GTIs) corresponding to the observational conditions using \texttt{maketime} and the \firstRev[minorRev]{extended housekeeping (EHK)} file of the target observation. We then extracted background events from the NXB database (v3) using \texttt{extractor}, and applied additional event screening with \texttt{ftselect}. Using the event files obtained after these data reduction steps, we generated NXB validation data with \texttt{rslnxbgen}. The NXB spectra were examined, and the model parameters were determined using C-statistics within \texttt{XSPEC} \citep{1996ASPC..101...17A}.

\subsection{Whole field-of-view spectral analysis}

For the observational data, we generated a pixel-by-pixel exposure map using \texttt{xaexpmap}, with the EHK file and the event file as inputs. We then created a redistribution matrix file (RMF) using \texttt{rslmkrmf}. Here, pixels 12 and 27 were excluded because they exhibit significantly degraded energy resolution and anomalous spectral responses compared to the nominal pixels. \firstRev[minorRev]{For the astrophysical emission, we used the ``L''-size Resolve RMF generated for the selected Hp events and retained pixels; this is distinct from the NXB RMF supplied with the version 3 recipe.} Similarly, an ancillary response file (ARF) excluding the same pixels was generated using \texttt{xaarfgen}.

For the ARF generation, \firstRev[minorRev]{an exposure-corrected Chandra image in the 2.0--10.0 keV band} was used as the input image representing the spatial distribution of photons centered on the pointing position. The Chandra data (ObsID 2220) were processed following standard data reduction procedures to produce the brightness map.

XRISM data reduction and screening were performed using \texttt{ftcopy} and \texttt{xselect}. In the analysis, pixels 12 and 27 were excluded, and only high-resolution primary events (``Hp'' or ITYPE = 0) were used to extract the image and the average spectrum.

The average spectrum was analyzed with \texttt{XSPEC}, and spectral fitting was performed using an absorbed \texttt{bapec} model (\texttt{phabs*bapec}). We adopted the protosolar abundance table of \cite{2009LanB...4B..712L}, implemented in XSPEC as \texttt{lpgs}. The hydrogen column density was fixed at the Galactic value in this direction, $N_{\mathrm{H}} = 4.9 \times 10^{20}\ \mathrm{cm}^{-2}$ \citep{2006PASJ...58..695H}. The NXB model constructed in the previous section was included as a fixed component in the fitting. The RMF and ARF described above were used, and the spectral fitting was carried out over the energy range of 2.0--10.0 keV. \firstRev[minorRev]{The best-fit parameters were obtained by minimizing the C-statistic.} Parameter uncertainties were evaluated using the \texttt{error} command in \texttt{XSPEC}, and are reported at the 90\% confidence level.

For comparison with other observations, a barycentric correction was applied using \texttt{barycen}. The Earth-centered relativistic Doppler velocity correction (ERADVEL) was $-19.978$ km~s$^{-1}$ and $-20.510$ km~s$^{-1}$ at the start and end of the observation, respectively. In this work, we adopt the average value of $-20.2$ km~s$^{-1}$. The redshift of the brightest cluster galaxy (BCG), NGC~3311, is $0.012759$ \citep{2011A&A...528A..24V}. Throughout this paper, all reported bulk velocities are barycentric-corrected and presented as relative velocities with respect to the BCG.

\subsection{Spatially resolved spectral analysis}

To investigate the spatial distribution of the spectra in more detail, we performed spatially resolved spectral analysis by dividing the field of view (FoV) into multiple regions. In this study, in addition to a method in which the spectra in each region are analyzed independently (non-SSM), we also carried out Spatial Spectral Mixing (SSM) analysis, which accounts for photon mixing from both inside and outside the FoV \citep{2018PASJ...70....9H,2025PASJ...77.1278X}.

\subsubsection{Nine-region division (non-SSM)}

Within the Resolve FoV, the $6 \times 6$ pixel array was divided into nine subregions, and regular spectral analysis was performed for each subregion. For each subregion, RMFs and ARFs were generated individually using the same procedure as in the whole-field analysis.

For the NXB spectra, we adopted the reference NXB model determined in the previous subsection and scaled its normalization factor according to the number of Resolve pixels included in each region. The same spectral model and fitting procedure as used in the whole-field analysis were applied to the spectral fitting. Results are described in section~\ref{sec:Nine-region maps}.

\subsubsection{Three-region division (SSM)}

Based on the results of the nine-region analysis, we divided the FoV into three regions to improve the statistical accuracy. In the SSM framework, it is essential to accurately reflect the surface brightness distribution of the source when generating ARFs to account for photon mixing. To this end, \firstRev[minorRev]{exposure-corrected Chandra images in the 2.0--10.0 keV band} corresponding to each subregion were constructed individually and used as brightness distribution models. Considering that the Resolve FoV is approximately 3 arcmin and that the half-power diameter of the point spread function (PSF) is 1.3 arcmin \citep{2025JATIS..11d2023I}, we defined regions outside the Resolve FoV within a radius of 3 arcmin from the pointing center as the ``out-of-field component'', and accounted for photon leakage from these regions.

\firstRev[minorRev]{The three in-field regions and the exterior brightness template used for the out-of-field component are illustrated in the right panel of figure~\ref{fig:rgb_combined_whiteDSS}. The latter represents sky emission outside the selected Resolve pixels, not an additional spectrum extracted from the detector. Its photons can enter the in-field extraction regions through the telescope PSF.}

\firstRev[minorRev]{In the simultaneous spectral fit, emission from the three in-field regions was propagated into each extracted spectrum through the appropriate cross-region ARFs. An additional \texttt{cflux*phabs*bapec} component describes out-of-field contamination in each spectrum. Its temperature, abundance, redshift, and velocity dispersion were tied to those of the corresponding in-field region. The in-field 2.0--10.0 keV fluxes were fitted, while the out-of-field flux was linked to the corresponding in-field flux using the surface-brightness ratio measured from the Chandra templates. Thus, the out-of-field flux was not an independent free parameter.}

\firstRev[minorRev]{This is a local spectral-template approximation, motivated by the relatively smooth thermodynamic structure of the core and the small out-of-field contribution (approximately 4--6\% in the ARF-based mixing fractions below). The Resolve data do not independently constrain the spectrum outside its FoV. We therefore do not infer a common physical velocity for all exterior gas from these ties; unmodeled spectral or velocity differences outside the FoV remain a systematic uncertainty.}

\section{Results}\label{sec:3}

In this section, we present the results of the spectral analysis of the central region of Abell~1060 obtained with XRISM/Resolve. We first describe the representative spectral properties of the ICM based on the full FoV spectrum. We then present the results of the spatially resolved analysis using the nine-region division to reveal the spatial distributions of physical quantities. Finally, we show the results of the three-region division analysis with the SSM, taking into account photon mixing from both inside and outside the FoV.

\subsection{FoV spectrum}

\begin{figure*}[htbp]
\begin{center}
\vspace{20pt}
\rotatebox{-90}{\includegraphics[width=5.5cm]{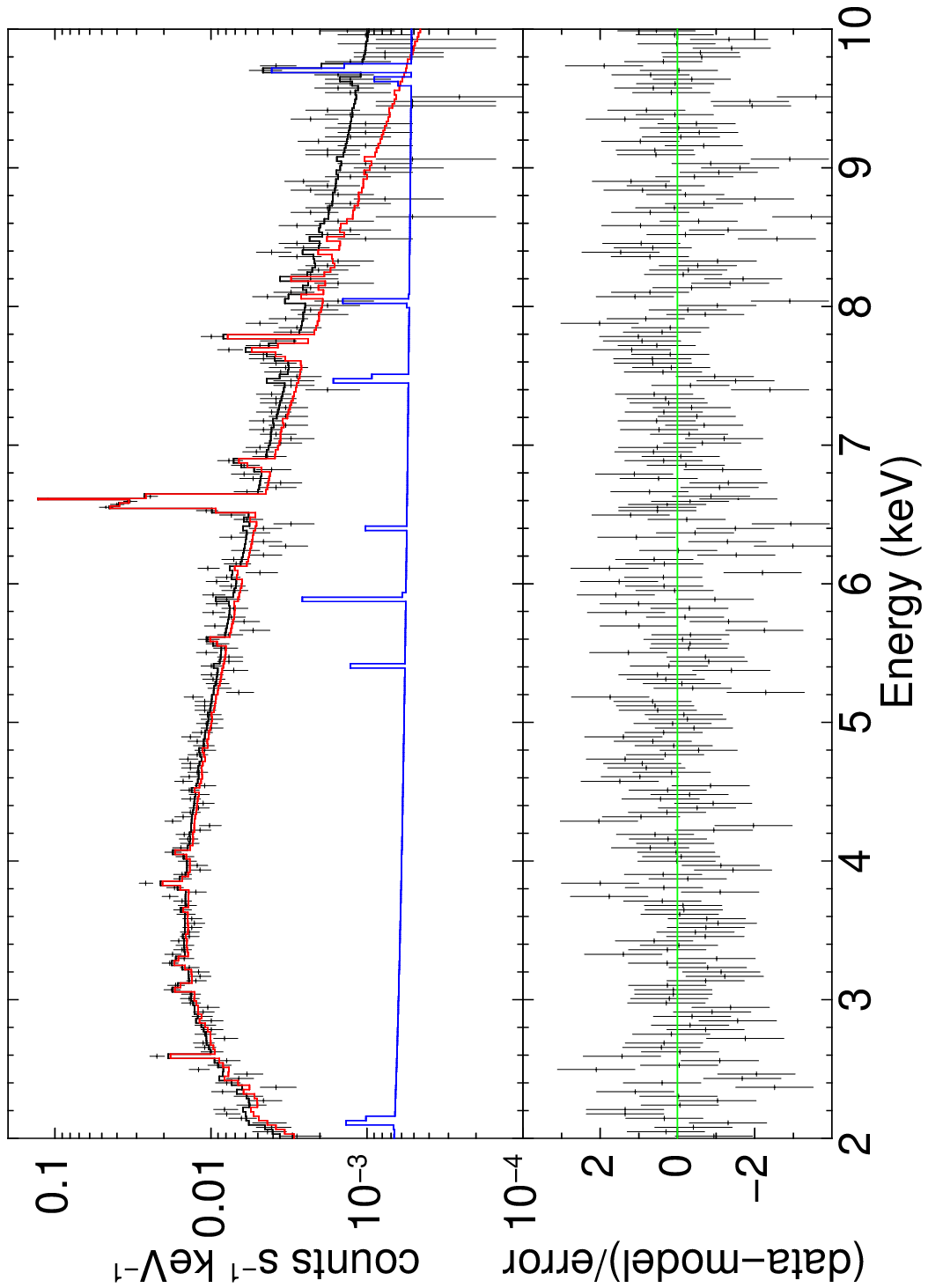}}
\rotatebox{-90}{\includegraphics[width=5.5cm]{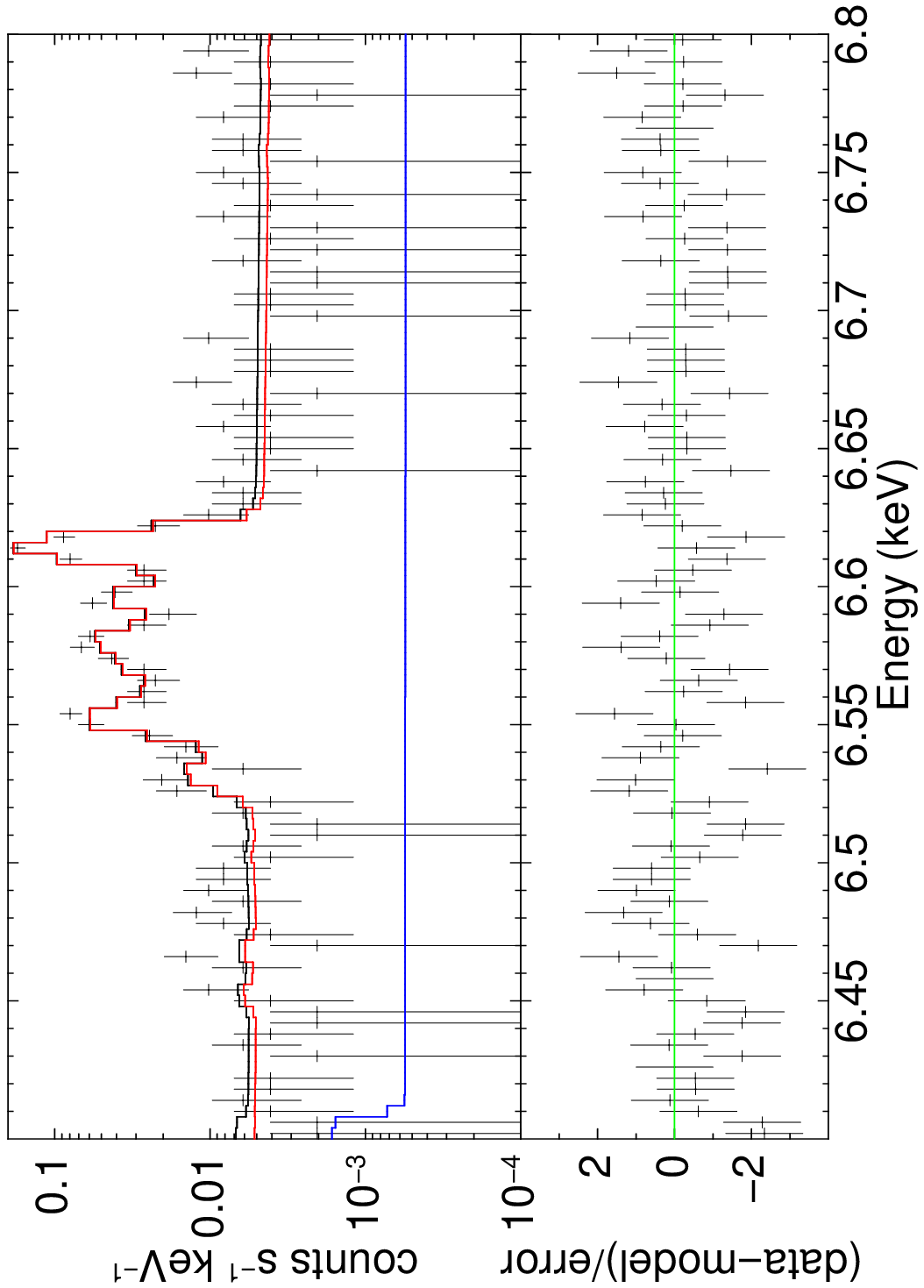}}
\end{center}
\vspace{-20pt}
\caption{
\firstRev[minorRev]{XRISM/Resolve spectrum of the full field of view (FoV) covering the central region of Abell~1060.}
(Left) Observed spectrum (black crosses) and the best-fit model (solid lines) in the 2.0--10.0 keV energy band. The \texttt{bapec} model is shown in red, the non-X-ray background (NXB) component in blue, and their sum in black. The lower panels show residuals between the data and the model, normalized by the statistical uncertainties. 
(Right) Same as the left panel, but zoomed into the Fe line energy band of 6.4--6.8 keV.\\
{Alt text: Two-panel XRISM Resolve spectrum. The left panel shows the full 2.0--10.0 keV spectrum and best-fit source and background models, while the right panel enlarges the 6.4--6.8 keV iron-line band. The residuals show no prominent systematic deviations.}
}
\label{fig:FoV_spectrum}
\end{figure*}

Figure~\ref{fig:FoV_spectrum} shows the results of the spectral fitting to the full FoV spectrum. The best-fit model reproduces the observed spectrum well, and no significant systematic residuals are found. The average temperature of the ICM is $3.32^{+0.13}_{-0.12}$ keV, the abundance is $0.65^{+0.06}_{-0.06}\ Z_\odot$, and the redshift is $(1.298^{+0.006}_{-0.006}) \times 10^{-2}$, all of which are consistent with previous studies \citep{2006PASJ...58..695H}. This measured redshift corresponds to a bulk velocity of $+45^{+18}_{-18}$ km~s$^{-1}$ relative to the BCG. \firstRev[minorRev]{Table}~\ref{tab:best_fit_parameter} summarizes the best-fit parameters of the \texttt{bapec} model.

\begin{table}[htbp]
\caption{Best-fit parameters with 90\% confidence errors.}\label{tab:best_fit_parameter}
\begin{center}
\begin{tabular}{lcc}
\hline
Parameter & Value & Unit \\
\hline
$kT$ & $3.32^{+0.13}_{-0.12}$ & keV \\
Abundance (lpgs) & $0.65^{+0.06}_{-0.06}$ & Z$_\odot$ \\
Bulk velocity & $+45^{+18}_{-18}$ & km~s$^{-1}$ \\
Velocity dispersion & $96.7^{+23.9}_{-24.6}$ & km~s$^{-1}$ \\
norm & $2.07^{+0.10}_{-0.09}\times10^{-2}{}$ & \firstRev[minorRev]{cm$^{-5}$} \\
\hline
\end{tabular}
\end{center}
\firstRev[minorRev]{The \texttt{bapec} normalization is $10^{-14}\int n_e n_{\rm H}\,dV/[4\pi\{D_A(1+z)\}^2]$, where $D_A$ is the angular-diameter distance in cm, $n_e$ and $n_{\rm H}$ are number densities in cm$^{-3}$, and $V$ is the emitting volume in cm$^3$.}
\firstRev[minorRev]{Abundances are relative to the protosolar values of \citet{2009LanB...4B..712L}, implemented as \texttt{lpgs} in XSPEC.}
\end{table}

\subsection{Nine-region maps}
\label{sec:Nine-region maps}

\begin{figure*}[h]
\begin{center}
\includegraphics[width=0.85\linewidth]{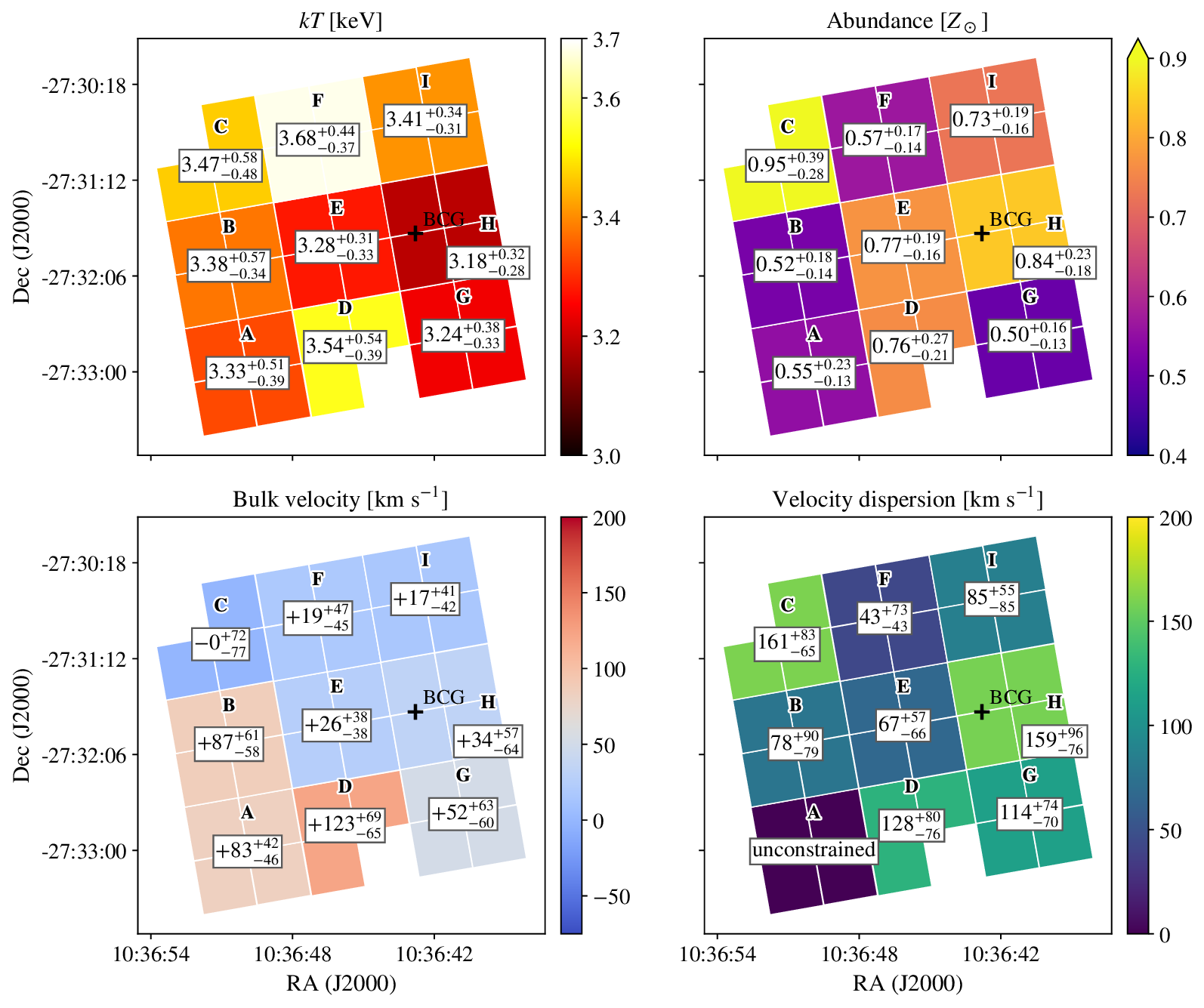}
\end{center}
\caption{
Spatial distributions obtained from the nine-region spatially resolved spectral analysis of the central region of Abell~1060 with XRISM/Resolve. Each colored square corresponds to a Resolve pixel. The top-left, top-right, bottom-left, and bottom-right panels show the temperature, abundance, line-of-sight velocity relative to the BCG, and velocity dispersion, respectively. \firstRev[minorRev]{The labels A--I identify the extraction regions used in Appendix~\ref{sec:parameter_profiles}. The values printed in each region are} the best-fit values with their 90\% confidence intervals. The black cross marks the center of the BCG (NGC~3311). \firstRev[minorRev]{The same values and uncertainties are shown as region-by-region plots in Appendix~\ref{sec:parameter_profiles} (figure~\ref{fig:parameter_profiles}) for improved readability.}\\
{Alt text: Four maps of the nine Resolve regions showing temperature, abundance, line-of-sight velocity, and velocity dispersion. The extraction regions are labeled A--I. Best-fit values and uncertainties are printed in each region, and a black cross marks NGC 3311. The velocity map shows a northwest-to-southeast gradient.}
}
\label{fig:9region_maps}
\end{figure*}

Figure~\ref{fig:9region_maps} shows the spatial distributions of temperature, abundance, velocity relative to the BCG, and velocity dispersion obtained from the spatially resolved spectral analysis with the $3 \times 3$ (nine-region) division of the FoV. \firstRev[minorRev]{For easier comparison of the values and their uncertainties, the same measurements are also shown as one-dimensional, region-by-region plots in Appendix~\ref{sec:parameter_profiles} (figure~\ref{fig:parameter_profiles}).}

Due to limited photon statistics in each subregion, statistically significant differences in these physical quantities cannot be robustly established. However, we focus on the overall spatial trends. While the temperature, abundance, and velocity dispersion do not exhibit significant spatial variations within the FoV, the velocity distribution appears to show a gradient across the regions.

\subsection{Three-region maps}
\label{sec:Three-region maps}

\begin{figure*}[h]
\begin{center}
\includegraphics[width=0.45\linewidth]{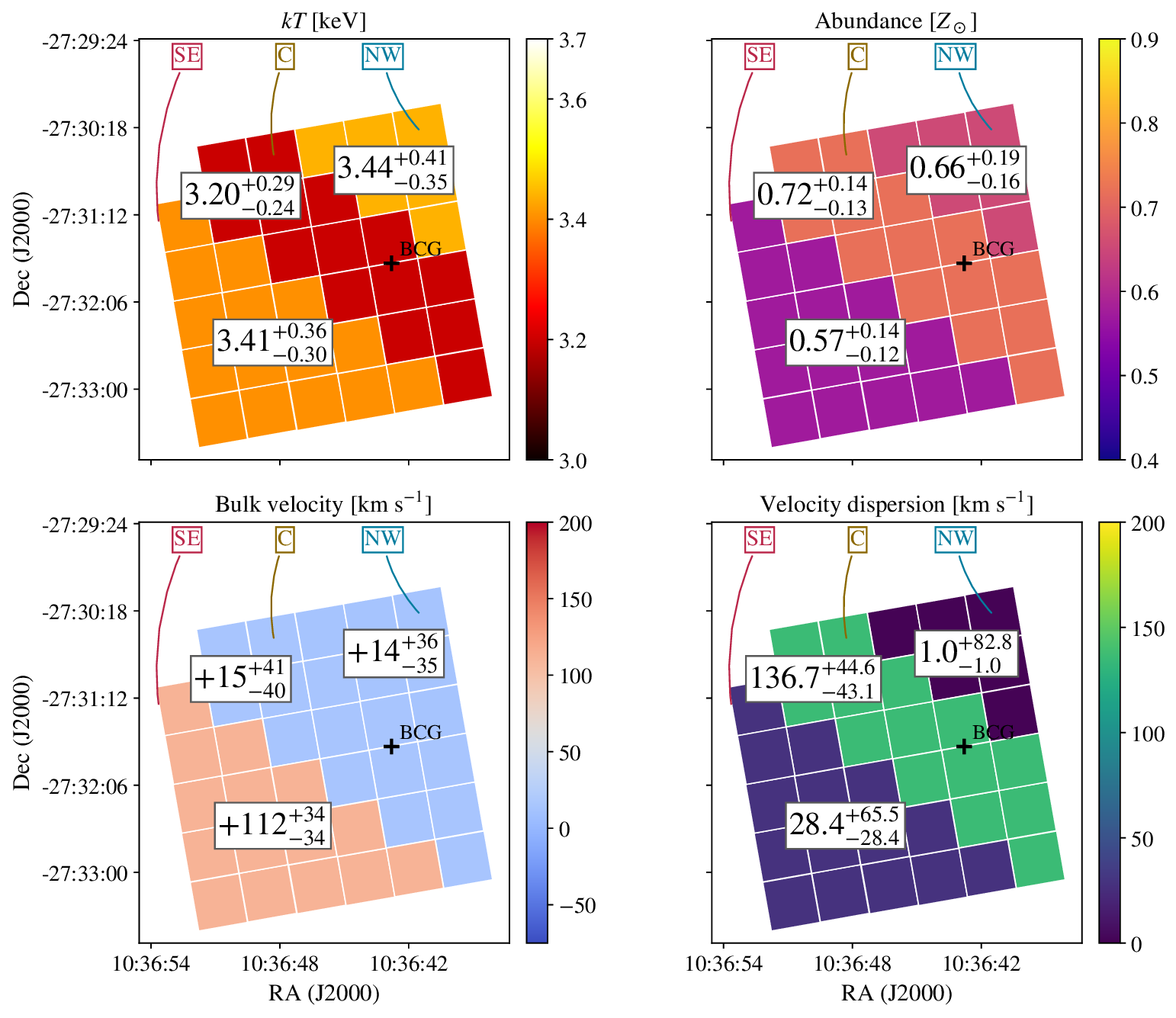}
\includegraphics[width=0.49\linewidth]{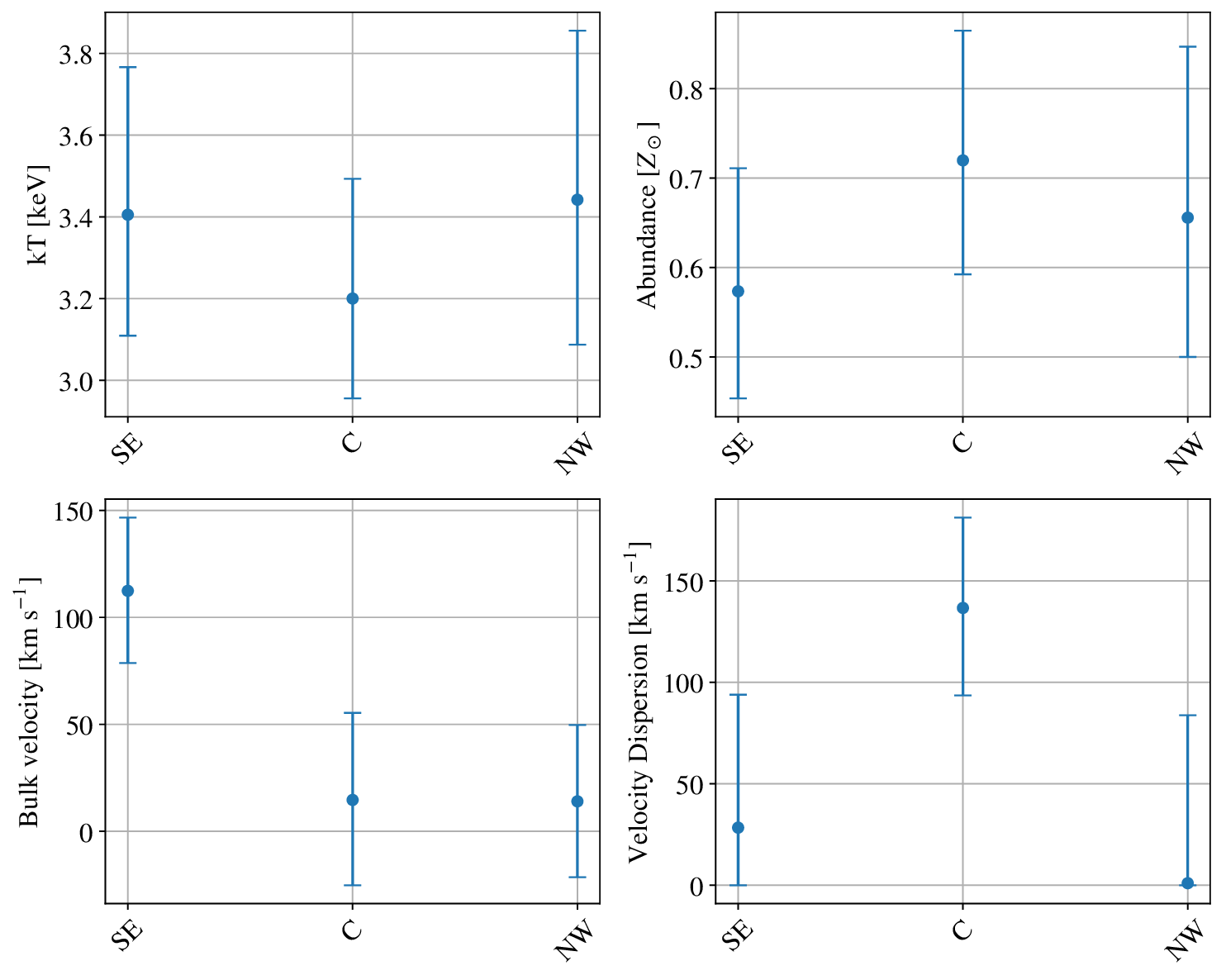}
\end{center}
\caption{
(Left) Spatial distributions of the temperature, abundance, line-of-sight velocity relative to the BCG, and velocity dispersion obtained for the three regions SE, C, and NW. The panel arrangement is the same as in \firstRev[minorRev]{figure}~\ref{fig:9region_maps}; the printed values are the best-fit values with their 90\% confidence intervals, and the black cross marks the center of the BCG (NGC~3311).
(Right) The same parameters plotted as a function of region, with error bars indicating the 90\% confidence intervals.\\
{Alt text: Comparison of physical parameters in the SE, C, and NW regions. The left panels map temperature, abundance, line-of-sight velocity, and velocity dispersion; the right panels plot the same measurements with error bars. The SE region has the largest positive velocity, while the central region has the largest velocity dispersion.}
}
\label{fig:ssm_maps}
\end{figure*}

We performed spatially resolved spectral analysis with a three-region division using SSM analysis to achieve higher statistical accuracy. The regions were defined along the northwest--southeast direction, where a possible velocity gradient was observed in \firstRev[minorRev]{figure}~\ref{fig:9region_maps}. \firstRev[minorRev]{Neighboring pixels were combined to obtain photon counts of the same order of magnitude over the full 2.0--10.0 keV fitting band, rather than specifically in the Fe-K band, while improving the statistics relative to the nine-region division.}

Figure~\ref{fig:ssm_maps} shows the spatial distributions of temperature, abundance, velocity relative to the BCG, and velocity dispersion. The figure also includes plots of these quantities as a function of region, where the region labels are shown on the horizontal axis and the corresponding values are plotted on the vertical axis. In this analysis, we simultaneously fitted a four-region model consisting of three in-field regions and one out-of-field component, thereby properly accounting for photon contributions from different regions. The best-fit parameters are summarized in \firstRev[minorRev]{table}~\ref{tab:ssm_result}.

With improved photon statistics in the three-region division, the spectral fits are well constrained and reproduce the data accurately. As detailed in Appendix~\ref{sec:appendix_ssm_spectra}, the dominant contribution to each spectrum arises from the corresponding region itself. To quantify the spectral mixing between regions, we evaluated the cross-region ARFs generated in the SSM analysis. For the NW region, the contributions from the NW, Center, SE, and out-of-field regions are \firstRev[minorRev]{79\%, 14\%, 1\%, and 6\%}, respectively. For the Center region, the corresponding contributions are \firstRev[minorRev]{25\%, 57\%, 13\%, and 4\%}, while for the SE region they are \firstRev[minorRev]{3\%, 21\%, 71\%, and 5\%}. Note that the fractional values do not sum exactly to 100\% due to rounding \firstRev[minorRev]{to the nearest percentage point}.


Significant variations in velocity are observed across the FoV. The velocities relative to the BCG are \firstRev[minorRev]{$+14$, $+15$, and $+112$ km~s$^{-1}$ in the NW, Center, and SE regions, respectively}. Although all regions are redshifted, the SE region exhibits a larger line-of-sight velocity relative to the others.

In contrast to the velocity structure, the velocity dispersion is found to be marginally larger in the central region than in the other two regions. This central region includes the BCG, NGC~3311. \firstRev[minorRev]{The central region also differs from the surrounding regions in its best-fit temperature and abundance, with a lower temperature and a higher abundance; however, these differences are not statistically significant when uncertainties are taken into account.}

\begin{table*}[htbp]
\caption{Best-fit parameters for the three regions with 90\% confidence errors.}
\label{tab:ssm_result}

\begin{center}
\begin{tabular}{lcccc}
\hline
& \multicolumn{3}{c}{Value} & \\
\cline{2-4}
Parameter &  NW & Center & SE & Unit \\

$kT$
& \firstRev[minorRev]{$3.44^{+0.41}_{-0.35}$}
& \firstRev[minorRev]{$3.20^{+0.29}_{-0.24}$}
& \firstRev[minorRev]{$3.41^{+0.36}_{-0.30}$}
& keV \\

Abundance (lpgs)
& \firstRev[minorRev]{$0.66^{+0.19}_{-0.16}$}
& \firstRev[minorRev]{$0.72^{+0.14}_{-0.13}$}
& \firstRev[minorRev]{$0.57^{+0.14}_{-0.12}$}
& Z$_\odot$ \\


Bulk velocity
& \firstRev[minorRev]{$+14^{+36}_{-35}$}
& \firstRev[minorRev]{$+15^{+41}_{-40}$}
& \firstRev[minorRev]{$+112^{+34}_{-34}$}
& ${\rm km~s^{-1}}$ \\

Velocity dispersion
& \firstRev[minorRev]{$1.0^{+82.8}_{-1.0}$}
& \firstRev[minorRev]{$136.7^{+44.6}_{-43.1}$}
& \firstRev[minorRev]{$28.4^{+65.5}_{-28.4}$}
& km~s$^{-1}$ \\

\firstRev[minorRev]{$\log_{10}[F_{2\text{--}10\,{\rm keV}}/({\rm erg\,cm^{-2}\,s^{-1}})]$}
& \firstRev[minorRev]{$-11.72^{+0.04}_{-0.04}$}
& \firstRev[minorRev]{$-11.33^{+0.03}_{-0.03}$}
& \firstRev[minorRev]{$-11.46^{+0.03}_{-0.04}$}
& --- \\

\hline
\end{tabular}
\end{center}

\firstRev[minorRev]{Abundances use the XSPEC \texttt{lpgs} protosolar reference of \citet{2009LanB...4B..712L}. $F_{2\text{--}10\,{\rm keV}}$ is the 2.0--10.0 keV energy flux; the table lists its base-10 logarithm in units of erg~cm$^{-2}$~s$^{-1}$.}
\end{table*}

\section{Discussion}\label{sec:4}

Compared with radio halos, FF has a much smaller physical extent and a morphology suggestive of a connection with the BCG, motivating us to consider fossil plasma supplied by past AGN activity as the primary origin of the radio-emitting electrons. We therefore first discuss the dynamical state of the ICM and the connection of FF with the central galaxies, and then test whether the presently observed turbulence could instead generate the required cosmic-ray electrons and magnetic field through in situ reacceleration and dynamo amplification.

\subsection{Dynamics of the ICM}

The observed velocity structure directly probes the dynamical state of the ICM and can reveal how fossil nonthermal plasma has been transported and reshaped. High-precision velocity measurements with XRISM therefore provide an important test of the proposed connection between gas motions, the central galaxies, and FF.

\subsubsection{Gas motions in Abell~1060}

As shown in \firstRev[minorRev]{section} 3.3, a significant velocity gradient was detected in the ICM. Interestingly, the direction of this velocity gradient is aligned with the elongation of FF observed in the radio. This suggests a possible connection between the large-scale gas motions in the ICM and the spatial distribution of the nonthermal emission. Such velocity structures may be associated with large-scale gas motions in the cluster core, such as sloshing, rotational motions, or residual flows induced by past merger events. In particular, the presence of a systematic velocity gradient cannot be explained solely by isotropic turbulence and instead indicates the contribution of more coherent bulk motions.

We further estimate the expected velocity perturbation induced by a passing massive perturber using the impulse approximation. Assuming a galaxy with a total mass of $M_{\rm pert} = 1.0 \times 10^{12}\ M_\odot$, a relative velocity of $V_{\rm rel} = 1000$ km~s$^{-1}$, and an impact parameter of $b = 100$ kpc---motivated by the spatial scale of the observed gradient---the induced velocity change caused by tidal force is $\Delta v \sim \frac{2 G M_{\rm pert}}{b V_{\rm rel}} \simeq 86$ km~s$^{-1}$. This value is comparable to the bulk velocity difference observed in this study ($\sim 100$ km~s$^{-1}$), indicating that the passage of a typical massive galaxy or a small subhalo could provide sufficient perturbation to generate the observed velocity structure.

\firstRev{NGC~3309 is one possible gravitational perturber, rather than a uniquely favored candidate. It is a giant elliptical galaxy projected close to NGC~3311, west of the BCG (figure~\ref{fig:rgb_combined_whiteDSS}). Its compact X-ray atmosphere, interpreted as being confined by the surrounding ICM pressure, supports its physical location within the dense cluster core \citep{2002ApJ...578..833Y,2011A&A...528A..24V}. The heliocentric recession velocities quoted by \citet{2011A&A...528A..24V} are $4099$ and $3825~{\rm km~s^{-1}}$ for NGC~3309 and NGC~3311, respectively, corresponding to $z\simeq0.01367$ and $0.01276$ in the optical $cz$ convention. Thus, NGC~3309 is redshifted relative to NGC~3311. These properties motivate considering NGC~3309 as a nearby galaxy that could perturb the central ICM, beyond the generic mass-scale argument alone.}

\firstRev{The XRISM velocity field shows that the SE gas is redshifted relative to the NW gas, although this gradient does not uniquely establish gas rotation. One possible encounter scenario is that a perturber passed on the southeastern side of the cluster centre with a line-of-sight velocity component directed away from us and excited the observed gas motions. The sign of the present velocity offset of NGC~3309 is compatible with this picture. A subsequent westward displacement along a non-radial orbit in the combined galaxy and cluster potential could, in principle, place the galaxy at its current projected position. 
We note that this proposed encounter scenario neither establishes a specific orbit nor constrains the time since the encounter. Likewise, the total mass of $10^{12}\,M_\odot$ adopted in the impulse estimate is an illustrative assumption rather than a measured mass of NGC~3309.}

Similarly, assuming a galaxy-group-scale perturber with $M_{\rm pert} = 10^{13}\ M_\odot$, a comparable velocity change would require an impact parameter of $b \sim 800$ kpc. This suggests that even large-scale substructures that do not directly pass through the cluster core may induce the observed velocity gradient through long-range gravitational perturbations. However, the FoV of XRISM/Resolve used in this study is limited to the central region of the cluster, preventing a direct constraint on the global velocity field at larger radii. In addition, wide-field observations with eROSITA do not show clear signatures of recent major merger events \citep{2026A&A...711A.226S}.

\subsubsection{Interaction with BCG}

In the central region of this cluster, the bright elliptical galaxy NGC~3311 is located and has been extensively investigated through optical observations \citep[e.g.,][]{2023A&A...679A..69I, 2024A&A...683A...8G}. In addition, previous observations have suggested the presence of galaxy motions \citep{2016A&A...589A.139B, 2018A&A...619A..70H} and intra-cluster light (ICL) \citep{2012A&A...545A..37A, 2018A&A...609A..78B}, indicating that interactions between galaxies and the ICM may not be negligible in the vicinity of the central galaxy.

The three-region analysis shows that the velocity dispersion in the region including NGC~3311 is marginally higher than in the surrounding regions, even when considering the 90\% confidence intervals. Such localized enhancements in velocity dispersion may correspond to turbulence or gas stirring induced by interactions between galaxies and the ICM, as has been observed in other clusters with XRISM \citep[e.g.,][]{2026ApJ...998..210X}. \firstRev[minorRev]{In addition, \citet{2026A&A...711A.226S} reported a $\Sigma$-shaped structure in an unsharp-masked Chandra image, associated with surface-brightness edges around the stripped halo of NGC~3311. This provides complementary morphological evidence for galaxy--ICM interactions, although it does not by itself identify the origin of the measured line broadening.}

In particular, localized turbulence driven by galaxy-scale processes around NGC~3311 may be present and observed as an enhancement in velocity dispersion. These results indicate that, in addition to the large-scale gas motions discussed in the previous subsection, galaxy-scale interactions also play an important role in shaping the dynamical state and turbulent structure of the ICM.

\subsection{Origin of FF}

Giant radio halos extend over several hundred kiloparsecs to megaparsec scales, much farther than synchrotron-emitting electrons can propagate within their radiative lifetimes, and therefore require in situ generation or reacceleration of cosmic-ray electrons \citep{brunetti2014cosmic}. In contrast, FF has a projected size of only $\sim30\times50$ kpc and lies close to the BCG, so the slow-diffusion problem that motivates reacceleration models for radio halos is not necessarily severe in this source.

The morphology and location of FF relative to NGC~3311 are therefore more naturally explained if the emitting particles originated in aged radio lobes produced by past activity of the BCG. \firstRev{We propose that this fossil plasma was disturbed together with the surrounding ICM and transported by coherent gas motions. The observed line-of-sight velocity gradient supports the presence of such motions, but neither identifies the perturber nor directly measures the transverse transport velocity.} In this picture, weak turbulence may subsequently modify or partially reaccelerate the plasma, but is not required to generate the cosmic-ray electron population, which is naturally provided by the pre-existing fossil plasma.

This scenario is consistent with the observed velocity structure revealed by XRISM, as well as optical evidence for dynamical disturbances in the cluster core, such as the ICL, which suggests ongoing perturbations by the external gravitational field \citep{2012A&A...545A..37A, 2018A&A...609A..78B}. Taken together, these observations support a picture in which fossil relativistic plasma has been redistributed by ICM motions and subsequently modified by weak turbulence in the cluster core.

\firstRev{To illustrate the transport timescale, we adopt a fiducial velocity in the plane of the sky of $v_\perp=100~{\rm km~s^{-1}}$. Its order of magnitude is motivated by the measured line-of-sight velocity difference of approximately $100~{\rm km~s^{-1}}$ across the central ICM, but the two quantities are not equivalent: a spatial velocity difference is not the speed of a plasma parcel relative to its injection site. We do not assume isotropy for this coherent flow or deproject the measured difference into a three-dimensional velocity. For steady transport over a projected separation $l_{\rm proj}$, the illustrative timescale is}
\begin{equation}\label{eq:tadv}
\firstRev{t_{\rm adv}=\frac{l_{\rm proj}}{v_\perp}\simeq294~{\rm Myr}\left(\frac{l_{\rm proj}}{30~{\rm kpc}}\right)\left(\frac{v_\perp}{100~{\rm km~s^{-1}}}\right)^{-1}.}
\end{equation}
\firstRev{This illustrative advection timescale is of the same order as the radiative cooling time of approximately 210 Myr estimated in section~\ref{sec:particle_acceleration}. This comparison suggests that advection may contribute to the transport of the fossil plasma, although it does not establish that transport is completed within the cooling time. The adopted transverse speed is motivated by the observed line-of-sight velocity difference, but is not directly measured by XRISM. Projection, the past flow history, and radiative aging remain uncertainties.}

Objects of this kind may be preferentially detected at low radio frequencies because of their steep radio spectra and inefficient particle acceleration, characteristics that are consistent with those of FF, which exhibits a steep radio spectral index of approximately $-1.4$ \citep{2024PASJ...76L...8K}. In recent years, several similar diffuse radio sources have been reported in nearby galaxy groups and clusters \citep[e.g.,][]{2024ApJ...962...40S, 2026A&A...710A...1S}, suggesting that FF may belong to a broader population of faint, aged synchrotron sources.

This advection scenario involves large-scale ICM motions induced by external perturbations. Similar dynamical disturbances are observed in other systems, such as the Centaurus cluster, where they manifest as gas sloshing and prominent cold fronts \citep[e.g.,][]{2016MNRAS.457...82S}. In the case of Abell~1060, the absence of such clear cold fronts is likely due to the lack of a strong temperature contrast in its relatively isothermal central region. Nevertheless, the macroscopic ICM motions responsible for transporting the fossil relativistic plasma of FF are a direct consequence of the same underlying dynamical processes that perturb the cluster core.

\subsection{Particle acceleration and dynamo processes}\label{sec:particle_acceleration}

Although the compact size and association with the BCG favor an AGN origin, the same framework commonly applied to radio halos provides a useful quantitative test of whether the present ICM turbulence could generate or maintain the nonthermal components of FF. We therefore evaluate the turbulent energy and energy flux using the velocity dispersion measured with XRISM and assess the efficiencies of turbulence-driven particle acceleration and magnetic field amplification.

\subsubsection{Dynamo efficiency}

We first estimate the turbulent kinetic energy density, $\epsilon_{\rm turb}$, from the velocity dispersion measured with XRISM. This estimate is based on the mass density of the ICM, $\rho_{\rm ICM}$, and the turbulent velocity dispersion, $\sigma_{\rm turb}$. The velocity dispersion derived from the \texttt{bapec} model corresponds to the unresolved turbulent velocity component in addition to thermal and instrumental broadening. As shown in the Results section, the velocity dispersion over the FoV is $\sigma_v = 96.7^{+23.9}_{-24.6}~{\rm km\,s^{-1}}$. Assuming that this unresolved velocity broadening is predominantly caused by turbulent gas motions, we use $\sigma_v$ as a proxy for the turbulent velocity ($\sigma_{\rm turb}$) dispersion in the following analysis.

The central electron number density of Abell~1060 is $n_e = 11.7 \times 10^{-3}\ \mathrm{cm}^{-3}$ \citep{2006PASJ...58..695H}. Adopting typical values for the mean molecular weight per electron, $\mu_e = 1.17$, and the proton mass, $m_{\rm p} = 1.67 \times 10^{-24}$ g, and assuming isotropic turbulence such that the measured velocity dispersion corresponds to the one-dimensional turbulent velocity, the turbulent energy density is given by
\begin{equation}\label{eq:eta_B}
\begin{aligned}
\epsilon_{\rm turb}
&= \frac{1}{2} \rho ( \sqrt{3}\sigma_{\rm turb} )^2 \\
&= \frac{1}{2} \mu_e m_{\rm p} n_e ( \sqrt{3}\sigma_{\rm turb} )^2 \\
&= 3.21 \times 10^{-12}\ \mathrm{erg\ cm^{-3}}.
\end{aligned}
\end{equation}

The resulting turbulent energy density can be compared with the magnetic energy density, $\epsilon_{\rm B}$, in the region as
\begin{equation}\label{eq:magnetic_energy}
\frac{B^2}{8\pi} = \eta_{\rm B}\epsilon_{\rm turb} = \eta_{\rm B}Q_{\rm turb}t_{\rm eddy},
\end{equation}
where $\eta_{\rm B}$ represents the fraction of turbulent kinetic energy converted into magnetic energy per eddy turnover time \citep{2022SciA....8.7623B}. $Q_{\rm turb}$ and $t_{\rm eddy}$ represent the turbulent energy flux and an eddy turnover time, respectively. The magnetic field strength can be estimated under the assumption of energy equipartition between cosmic rays and magnetic fields. Assuming that the line-of-sight depth of FF is comparable to its projected size (i.e., $l \sim 50$ kpc), we estimate the equipartition magnetic field strength using the revised equipartition formula of \citet{2005AN....326..414B}. \firstRev[minorRev]{For comparison, \citet{2004MNRAS.352...76P} discuss proton-to-electron ratios of order 100--300 in hadronic models of Perseus and Coma. We use $K_0=100$ as a fiducial proton-to-electron number ratio, motivated by the lower end of these model estimates, not as a measurement of the particle content of FF. The equipartition estimate consequently assumes a proton-containing plasma; an aged AGN plasma may have a different and energy-dependent ratio.} Adopting $K_0 = 100$, a 330~MHz radio surface brightness of $I_{330\,{\rm MHz}} = 1.0$ mJy beam$^{-1}$, a synchrotron spectral index of $\alpha=-1.4$ \citep{2024PASJ...76L...8K}, a minimum cosmic-ray proton energy of $E_{\rm p,min}=938$ MeV, and an isotropic magnetic field, we obtain an equipartition magnetic field strength of $B_{\rm eq} \simeq 4.14~\mu$G. Note that if a different particle ratio is assumed, the estimated field strength scales weakly as $B_{\rm eq} \propto (1 + K_0)^{\frac{1}{3 - \alpha}}$.

\firstRev[minorRev]{At fixed $B_{\rm eq}$ and gas density, $\eta_{\rm B}\propto\sigma_v^{-2}$. Propagating the 90\% confidence interval $\sigma_v=72.1$--$120.6~{\rm km~s^{-1}}$ therefore gives $\eta_{\rm B}=0.21^{+0.17}_{-0.07}$ (approximately 0.14--0.38). This interval includes only the statistical uncertainty in the velocity dispersion, not the systematic uncertainties in the equipartition field, particle content, or gas density.} In other words, the magnetic energy density is approximately 21\% of the turbulent energy density. \firstRev[minorRev]{The central value is a factor of approximately 4--7 above the efficiencies} derived from magnetohydrodynamic (MHD) simulations, which typically find $\eta_{\rm B}\sim0.03$--0.05 \citep[e.g.,][]{Beresnyak_2012}. Equivalently, the turbulent energy density is approximately 4.8 times larger than the magnetic energy density. The velocity dispersion measured with XRISM reflects the present level of turbulence, whereas the magnetic field may retain information about the past dynamical evolution of the ICM. One possible explanation is that the turbulence in FF was stronger in the past and has subsequently decayed. \firstRev[minorRev]{These simulation efficiencies are reference values, rather than universal upper limits, and the comparison depends on the assumed magnetic field and on how well the measured velocity dispersion represents the turbulent energy.} Alternatively, the magnetic field may also have been amplified by other processes, such as AGN feedback or gas compression associated with flux freezing.

In addition, the velocity dispersion measured by XRISM does not directly constrain the turbulence at the driving scale that amplifies the magnetic field. If a significant fraction of the turbulent energy resides on scales that are not adequately captured by the observed velocity dispersion, the velocity dispersion will be underestimated, and hence the value of $\eta_{\rm B}$ will be overestimated. We also note that significant uncertainties are associated with the estimate of the magnetic field strength and the line-of-sight path length. \citet{2026A&A...705A.129V} suggested that the velocity dispersion measured with XRISM/Resolve may be systematically biased by emissivity weighting effects, which depend on the volume filling factor and thermodynamic structure of the emitting gas.

To illustrate the impact of these uncertainties, \firstRev[minorRev]{figure}~\ref{fig:eta_acc_B} shows the dependence of $\eta_{\rm B}$ on both the magnetic field strength and the turbulent velocity. As shown in \firstRev[minorRev]{equations~\ref{eq:eta_B} and~\ref{eq:magnetic_energy}}, $\eta_{\rm B}$ depends strongly on the turbulent velocity, scaling approximately as $\sigma_{\rm turb}^{-2}$, such that a factor of $\sim 3$ variation in velocity dispersion can lead to an order-of-magnitude change in the inferred dynamo efficiency. If the observed XRISM/Resolve velocity dispersion is representative of the true turbulent velocity, achieving a physically plausible value of $\eta_{\rm B}$ would favor a magnetic field strength somewhat lower than the equipartition estimate.

\subsubsection{Turbulent acceleration efficiency}

We next evaluate the efficiency of particle acceleration driven by turbulence. This efficiency can be characterized by the turbulent energy flux, $Q_{\rm turb}$, and the nonthermal emissivity, $Q_{\rm NT}$ (including synchrotron and inverse Compton emission). By comparing these quantities with the observed synchrotron luminosity, we assess the efficiency of turbulent acceleration, following the method of \cite{2022SciA....8.7623B}.

The turbulent energy flux is given by $Q_{\rm turb} = \epsilon_{\rm turb}/\tau_{\rm eddy}$, where $\tau_{\rm eddy}$ is the eddy turnover timescale. To estimate $\tau_{\rm eddy}$, it is necessary to assume the spatial scale associated with the measured velocity dispersion. \firstRev[minorRev]{Since the turbulence injection scale is not directly constrained by the XRISM observations,} we adopt the size of the XRISM/Resolve FoV, $L = 50$ kpc (see \firstRev[minorRev]{figure}~\ref{fig:rgb_combined_whiteDSS}), as the representative spatial scale of the unresolved turbulent motions. The corresponding eddy turnover time is $\tau_{\rm eddy} = L/(\sqrt{3}\sigma_{\rm turb}) = 292$ Myr, yielding $Q_{\rm turb} \approx 3.49 \times 10^{-28}\ {\rm erg\ cm^{-3}\ s^{-1}}$. Here, we assume that the velocity dispersion is dominated by unresolved turbulent eddies on scales $\lesssim L$ and that the largest scale within this unresolved range carries the dominant fraction of the turbulent energy.

We then estimate the nonthermal emissivity $Q_{\rm NT}$. The radio flux density of FF has been measured at low frequencies \citep{2024PASJ...76L...8K}, allowing an estimate of the monochromatic synchrotron luminosity as $\nu L_{\nu} = 4\pi D_{\rm L}^2 \nu F_{\nu} = 2.48 \times 10^{38}\ {\rm erg\ s^{-1}}$. Here, for simplicity, we assume that the radio flux density at 338 MHz represents the characteristic luminosity and neglect the integration over the full spectral shape. Including inverse Compton losses against the cosmic microwave background (CMB), the equivalent magnetic field strength is $B_{\rm CMB} = 3.25(1+z)^2 \approx 3.3\ \mu$G. The total nonthermal luminosity is given by $\nu L_{\rm NT} = \nu L_{\rm sync}(1 + B_{\rm CMB}^2/B^2)$, yielding $\nu L_{\rm NT} = 4.09 \times 10^{38}\ {\rm erg\ s^{-1}}$. 

Assuming that the projected size of FF is $\sim 30 \times 50$ kpc and that the line-of-sight depth is $\sim 50$ kpc, the emitting volume is $V \approx 7.5 \times 10^{4}$ kpc$^3$. The corresponding nonthermal emissivity is then $Q_{\rm NT} = \nu L_{\rm NT}/V \approx 1.86 \times 10^{-31}\ {\rm erg\ cm^{-3}\ s^{-1}}$.

From these estimates, we obtain an acceleration efficiency of $\eta_{\rm CRe} = Q_{\rm NT}/Q_{\rm turb} = 5.3 \times 10^{-4}$, indicating that approximately 0.05\% of the turbulent energy flux is converted into particle acceleration. However, this estimate is subject to large uncertainties, primarily due to the assumed spectral shape. The bolometric synchrotron luminosity depends sensitively on the lower cutoff of the electron Lorentz factor ($\gamma_{\rm min}$). For example, fixing $\gamma_{\rm max} = 10^4$, the bolometric luminosity becomes larger than the monochromatic estimate by factors of $\sim 50$, $\sim 13$, and $\sim 7$ for $\gamma_{\rm min} = 100, 500,$ and $1000$, respectively. Therefore, the monochromatic approximation used above can be regarded as a lower limit on the total nonthermal luminosity.

\begin{figure}[h]
\begin{center}
\includegraphics[width=\linewidth]{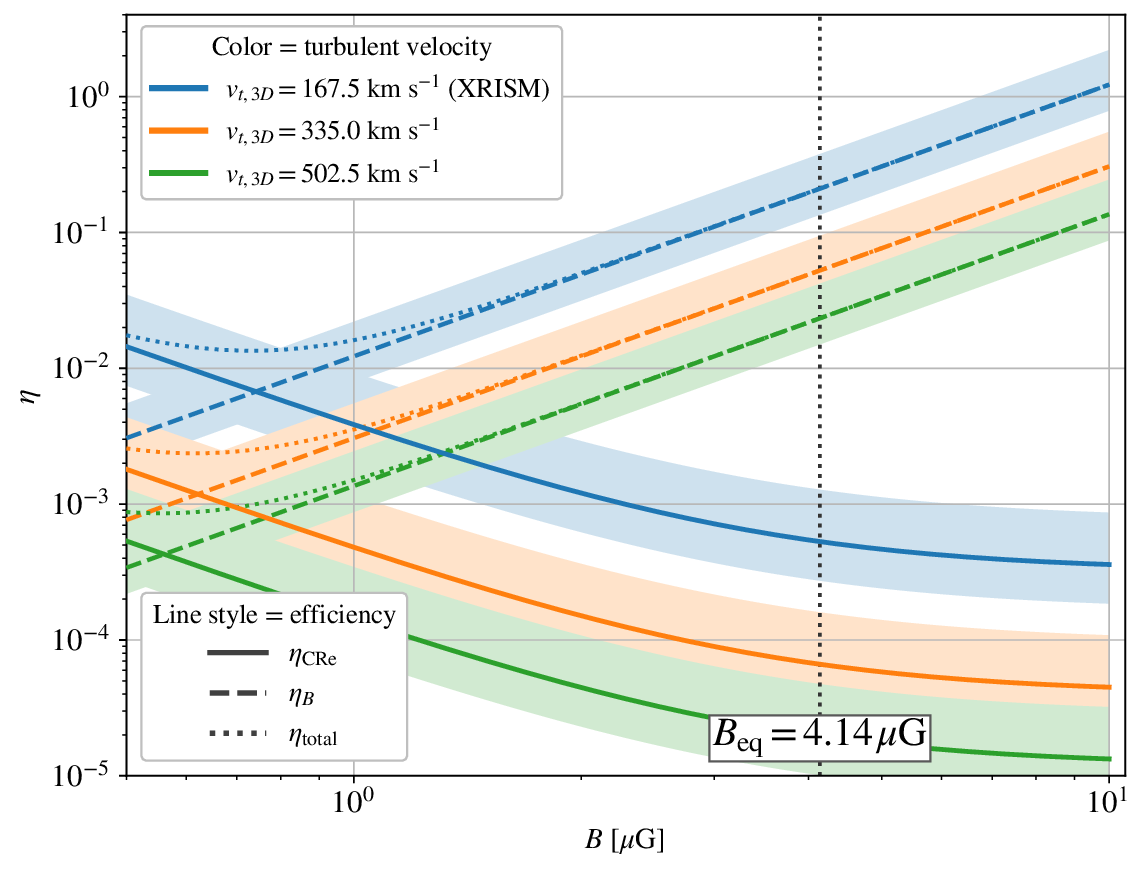}
\end{center}
\caption{
Distributions of the CRe acceleration efficiency ($\eta_{\rm CRe}$; solid lines), dynamo efficiency ($\eta_{\rm B}$; dashed lines), and their sum ($\eta_{\rm total}$; dotted lines) as functions of magnetic field strength. Blue, orange, and green indicate turbulent velocity dispersions of 1, 2, and 3 times the observed value, respectively. \firstRev[minorRev]{The pale bands show the 90\% ranges in both $\eta_{\rm CRe}$ and $\eta_{\rm B}$ obtained by propagating the velocity-dispersion uncertainties at fixed magnetic field and gas density.} The vertical dotted line marks the equipartition magnetic field, $B_{\rm eq}=4.14~\mu$G.\\
{Alt text: Logarithmic curves of cosmic-ray electron acceleration, dynamo, and total efficiencies versus magnetic field strength for turbulent velocities of one, two, and three times the observed value. \firstRev[minorRev]{Shaded uncertainty bands surround the acceleration and dynamo efficiency curves,} and a vertical line marks the equipartition field at 4.14 microgauss.}
}
\label{fig:eta_acc_B}
\end{figure}

The observationally estimated CRe acceleration efficiency ($\sim 0.05$--$2.5$\%) is consistent with the relatively low acceleration efficiencies predicted by recent cosmological simulations of galaxy clusters and numerical models \citep[e.g.,][]{2022SciA....8.7623B, 2024ApJ...961...15N, Beduzzi_2024}. This agreement shows that weak turbulent reenergization of a pre-existing electron population is energetically possible, but the efficiency alone does not demonstrate that the electrons were generated in situ. The acceleration-timescale test below provides a stronger constraint on that interpretation. Finally, as shown in \firstRev[minorRev]{figure}~\ref{fig:eta_acc_B}, the inferred efficiency depends sensitively on the assumed magnetic field strength and turbulence scale, both of which remain subject to significant uncertainties. For comparison, the efficiency distributions obtained under different assumed scales, while maintaining a constant velocity dispersion, are presented in Appendix \ref{sec:A-1}.

\subsubsection{Reacceleration timescale}

\firstRev[minorRev]{For turbulent reacceleration to sustain the synchrotron emission,} the acceleration timescale must be comparable to or shorter than the radiative cooling timescale. The turbulent acceleration timescale can be evaluated once the turbulence strength and spatial scale are specified. In the framework of the adiabatic stochastic acceleration (ASA) model, the reacceleration timescale is given by \citep{2016MNRAS.458.2584B}

\begin{equation}\label{eq:tacc}
\tau_{\rm acc}
\simeq
\frac{\sqrt{6/5}}{12}
\frac{c}{c_s^2}
\frac{L_0}{\sqrt{\beta_{\rm pl}}}
\,M_t^{-3}
\,\psi^{3},
\end{equation}

\noindent
where $c$ is the speed of light, $c_s$ is the sound speed in the ICM, $L_0$ is the turbulence injection scale, $\beta_{\rm pl}$ is the plasma beta, $M_t$ is the turbulent Mach number, and $\psi = \lambda_{\rm mfp}/l_A$ is the mean free path of cosmic-ray electrons normalized by the Alfv\'en scale.

Based on the observational results and the assumptions adopted in this study, we take $L_0 = 50$ kpc, $k_BT = 3.3$ keV, $B = 4.14\ \mu$G, $v_{\rm turb} = 96.7$ km~s$^{-1}$, $n_e = 11.7 \times 10^{-3}\ \mathrm{cm}^{-3}$, and $\psi = 0.5$. The sound speed in the ICM is given by $c_s = \sqrt{\gamma k_BT / \mu m_p}$, where $\gamma = 5/3$ is the adiabatic index, $k_B$ is Boltzmann's constant, $T$ is the temperature, and $\mu = 0.6$ is the mean molecular weight per particle. Substituting $kT = 3.3$ keV yields $c_s \simeq 940$ km~s$^{-1}$. The 3D turbulent Mach number is then $M_t = \sqrt{3}~v_{\rm turb}/c_s$, giving $M_t \simeq 0.18$. The plasma beta is given by $\beta_{\rm pl} = 8\pi n_{\rm tot} kT / B^2$, where the total particle number density is approximated as $n_{\rm tot} \simeq 1.8n_e$ for a plasma with cosmic abundances. We then obtain $\beta_{\rm pl} \simeq 163$. Substituting these values into \firstRev[minorRev]{equation}~(\ref{eq:tacc}) yields an acceleration timescale of $\tau_{\rm acc} \simeq 2.5$ Gyr.

Considering that the observed synchrotron spectrum is steep, the relativistic electrons are expected to suffer significant radiative cooling. We estimate the cooling timescale using the break frequency of the radio spectrum and compare it with the acceleration timescale. The Lorentz factor of cosmic-ray electrons emitting synchrotron radiation at a given observing frequency can be estimated from
\begin{equation}\label{eq:synchrotron_frequency}
\nu_{\rm syn} = 4.2 \, \gamma^2 B_{\mu{\rm G}} \ {\rm Hz},
\end{equation}

\noindent
where $B_{\mu{\rm G}}$ is the magnetic field strength in units of $\mu{\rm G}$ \citep{2020MNRAS.497.4704G}. Following \citet{1999ApJ...520..529S}, we convert the break frequency, $\nu_{\rm b}$, into the corresponding electron Lorentz factor and \firstRev{estimate the radiative cooling timescale from synchrotron and inverse Compton losses, $\tau_{\rm cool}=\gamma/(b_{\rm syn}+b_{\rm IC})$, with $b_{\rm syn}=1.30\times10^{-21}\gamma^2 B_{\mu{\rm G}}^2~{\rm s^{-1}}$ and $b_{\rm IC}=1.37\times10^{-20}\gamma^2(1+z)^4~{\rm s^{-1}}$. For $B=4.14~\mu$G and $\nu_{\rm b}=300$ MHz, this gives approximately 208 Myr, rounded to $\tau_{\rm cool}\simeq210$ Myr.}

\noindent

\begin{figure}[h]
\begin{center}
\includegraphics[width=\linewidth]{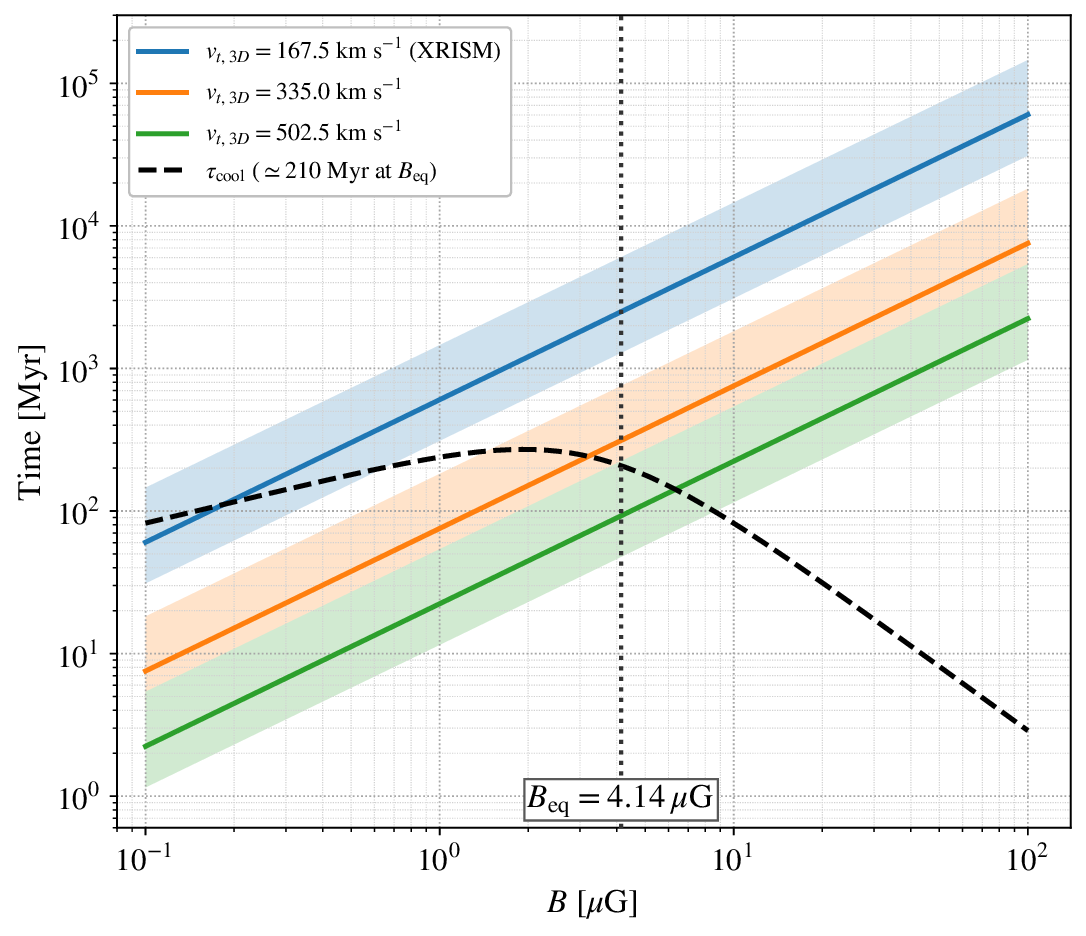}
\end{center}
\caption{
Acceleration timescale ($\tau_{\rm acc}$; colored solid lines) and radiative cooling timescale ($\tau_{\rm cool}$; black dashed line) as functions of magnetic field strength. Blue, orange, and green indicate turbulent velocity dispersions of 1, 2, and 3 times the observed value, respectively. \firstRev[minorRev]{The shaded bands show the ranges in $\tau_{\rm acc}$ propagated from the 90\% velocity-dispersion interval.} \firstRev{The cooling curve follows $\tau_{\rm cool}\propto B^{1/2}/(B^2+B_{\rm CMB}^2)$ and is calculated from synchrotron and inverse Compton losses at 300 MHz, giving approximately 210 Myr at $B_{\rm eq}$.} The vertical dotted line marks $B_{\rm eq}=4.14~\mu$G.
\\
{Alt text: Acceleration and radiative cooling timescales versus magnetic field strength. Colored curves show acceleration times for turbulent velocities of one, two, and three times the observed value, with shaded uncertainty bands; a black dashed curve shows the cooling time, and a vertical line marks 4.14 microgauss.}
}
\label{fig:tau_acc}
\end{figure}

\firstRev[minorRev]{Figure}~\ref{fig:tau_acc} shows the distribution of $\tau_{\rm acc}$ obtained by varying the magnetic field strength and turbulence strength. Under the representative parameters adopted in this study, the acceleration timescale is estimated to be $\tau_{\rm acc} \sim 2.5$ Gyr, which is much longer than the cooling timescale inferred from the synchrotron emission, $\tau_{\rm cool} \sim 210$ Myr. \firstRev[minorRev]{Since sustaining the synchrotron emission through turbulent reacceleration requires} $\tau_{\rm acc} \lesssim \tau_{\rm cool}$, this result suggests that the currently observed level of turbulence alone is insufficient to account for the acceleration of electrons. However, recent studies have pointed out that XRISM-like measurements may underestimate the true level of turbulent velocity, which could potentially alleviate this tension \citep{2026A&A...705A.129V}. Since the acceleration timescale depends strongly on the turbulent Mach number ($\tau_{\rm acc} \propto M_{\rm t}^{-3}$; \firstRev[minorRev]{equation}~\ref{eq:tacc}), modest systematic uncertainties in the measured velocity dispersion can have a substantial impact on the inferred reacceleration efficiency. As shown in \firstRev[minorRev]{figure}~\ref{fig:tau_acc}, a factor-of-two increase in the turbulent velocity would shorten the acceleration timescale by nearly an order of magnitude. Quantifying emissivity-weighting effects, therefore, remains essential, although this is challenging with the currently available XRISM data alone.

This result can be interpreted as indicating that the turbulent Mach number is too low for turbulent reacceleration to be effective. Under a higher turbulent Mach number, the acceleration timescale is significantly reduced according to $\tau_{\rm acc} \propto M_t^{-3}$, allowing electrons to be rapidly accelerated to high energies. The currently observed weaker turbulence may then play a role only in sustaining these electrons after the initial acceleration phase, rather than driving their acceleration.

These quantitative results reinforce the interpretation introduced above: the present turbulence is more naturally understood as a process that weakly reenergizes or maintains fossil electrons and modifies an existing magnetic field, rather than as the mechanism that created both components in situ. The long acceleration time and high inferred dynamo efficiency therefore support a scenario in which energetic electrons and magnetized plasma were supplied by past AGN activity and subsequently advected by ICM motions. This interpretation is also consistent with the steep radio spectral index of FF ($\alpha \simeq -1.4$), as weak reacceleration would not efficiently compensate for the radiative aging of the electron population.

\section{Summary}\label{sec:5}

In this study, we used high-resolution spectroscopy with XRISM/Resolve to measure the velocity structure and velocity dispersion in the central region of Abell~1060 and investigate the origin of FF. We first examined the connection between FF and the dynamical state of the ICM, and then quantitatively tested whether the presently observed turbulence could generate or maintain its nonthermal components. The main results and interpretations are summarized as follows.

Spatially resolved X-ray spectroscopy reveals a systematic velocity gradient in the ICM whose direction is aligned with the elongation of FF. This indicates that coherent, large-scale gas motions play an important role in shaping the spatial distribution of the nonthermal plasma. Estimates based on the impulse approximation show that gravitational perturbations from galaxy- or group-scale objects can plausibly produce the observed velocity structure.

We also find a localized enhancement in velocity dispersion near the BCG, NGC~3311, suggesting gas stirring or turbulence induced by galaxy--ICM interactions. Together with the morphology and compact physical extent of FF, these kinematic features support a scenario in which fossil plasma injected by past BCG activity was advected and reshaped by ICM motions. \firstRev{For a fiducial transverse advection velocity of $100~{\rm km~s^{-1}}$, the transport time over the projected distance of $\sim30$ kpc is approximately $300$ Myr, of the same order as the estimated radiative cooling time of $\sim210$ Myr. This comparison suggests that advection may contribute to the transport of fossil plasma, but does not establish that transport is completed within the cooling time. This is a conditional timescale estimate: XRISM measures line-of-sight velocities, not the transverse transport speed.}

The velocity dispersion measured over the Resolve FoV is $96.7^{+23.9}_{-24.6}$ km~s$^{-1}$. The corresponding turbulent energy flux is sufficient to weakly reenergize a pre-existing population of relativistic electrons. However, the magnetic field amplification efficiency inferred from the same turbulence level exceeds typical theoretical and numerical predictions, suggesting that the presently observed turbulence alone is unlikely to have generated the magnetic field of FF.

The turbulent acceleration timescale is estimated to be $\sim2.5$ Gyr, substantially longer than the radiative cooling timescale of $\sim210$ Myr. Thus, the present turbulence is insufficient to generate the radio-emitting electrons in situ, although weak reacceleration may help maintain electrons previously supplied by AGN activity. Within the turbulent reacceleration framework, a turbulent velocity at $L=50$ kpc approximately 2.3 times the value inferred from XRISM would be required to make the acceleration and cooling timescales comparable. This limited role of the present turbulence is also consistent with the steep radio spectral index of FF ($\alpha\simeq-1.4$).

Taken together, these results favor a scenario in which magnetized plasma and relativistic electrons produced by past AGN activity are transported by cluster-scale gas motions and subsequently modified by weak turbulence. This study demonstrates that XRISM velocity measurements can directly connect the dynamical history of the ICM to the origin and evolution of nonthermal structures. Future wide-field and multiwavelength observations with facilities such as SKA and ngVLA will further test this picture.


\section*{\firstRev[minorRev]{Funding}}
\firstRev[minorRev]{This work was supported by JSPS KAKENHI Grant Number 25KJ0162 (KK).}

\begin{ack}
This paper employs a list of Chandra datasets, obtained by the Chandra X-ray Observatory, contained in the Chandra Data Collection (CDC). 
We thank the XRISM team for their contributions to the development and operation of the mission and its data-analysis software.
\end{ack}

\appendix
\section{\firstRev[minorRev]{Region-by-region spectral parameters}}
\label{sec:parameter_profiles}
\begin{figure*}[htbp]
\centering
\includegraphics[height=0.62\textheight]{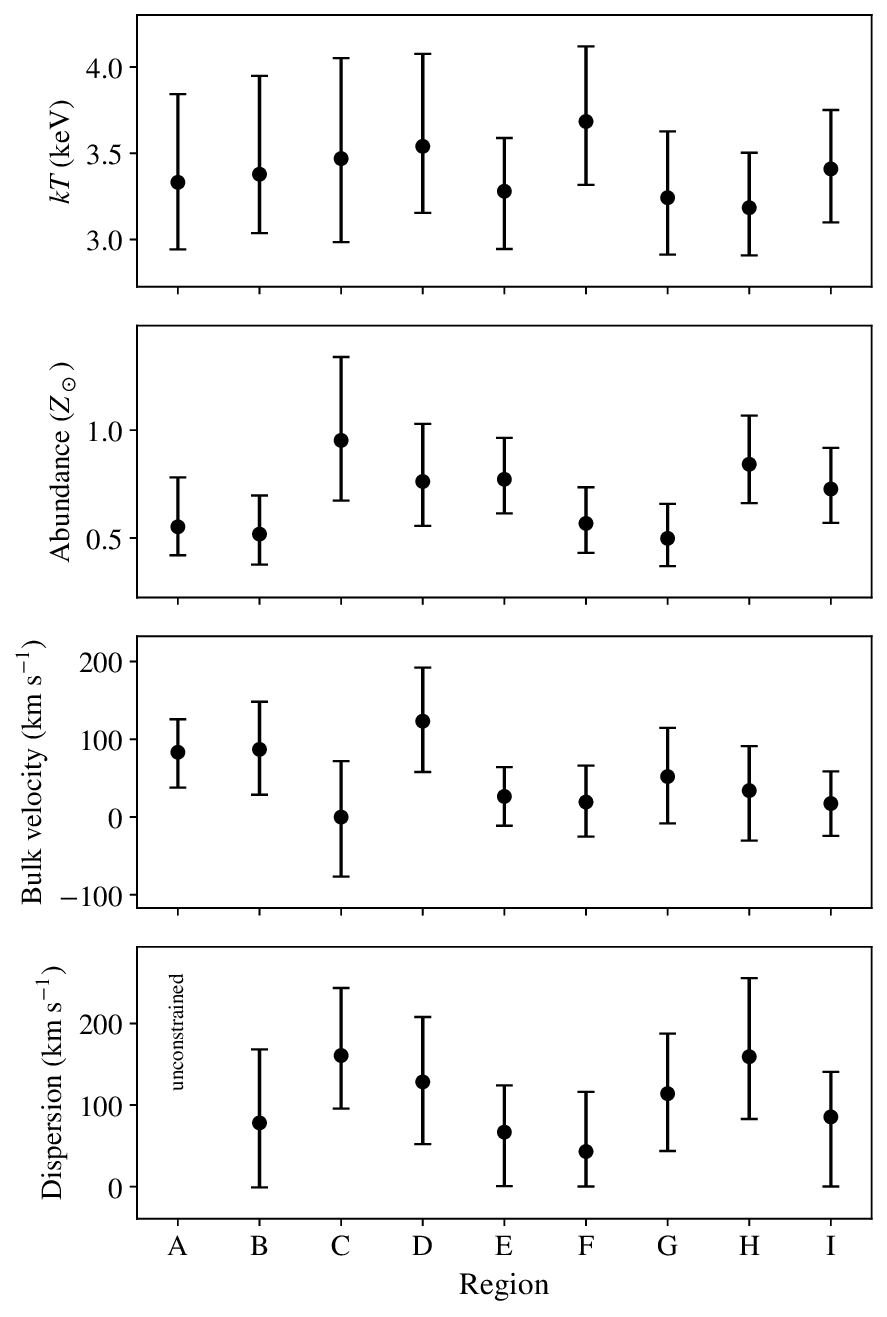}
\caption{\firstRev[minorRev]{Temperature, abundance, line-of-sight velocity relative to the BCG, and velocity dispersion for the nine-region analysis. Error bars show 90\% confidence intervals. The measurements are the same as those in the main-text maps; no new fits are performed.}\\
\firstRev[minorRev]{Alt text: Four error-bar plots comparing spectral parameters across nine regions A--I. The velocity dispersion of region A is labeled unconstrained.}}
\label{fig:parameter_profiles}
\end{figure*}

\firstRev[minorRev]{Figure~\ref{fig:parameter_profiles} presents the measurements shown in figure~\ref{fig:9region_maps} as region-by-region plots. These are categorical comparisons, not radial profiles. The nine-region labels A--I follow the extraction-region identifiers. Region A has no constrained velocity-dispersion interval in the saved fit results and is therefore labeled as unconstrained rather than assigned an error bar.}

\section{Spectra of the three-region SSM analysis}
\label{sec:appendix_ssm_spectra}

\begin{figure}[h]
\begin{center}
\rotatebox{-90}{\includegraphics[width=5.5cm]{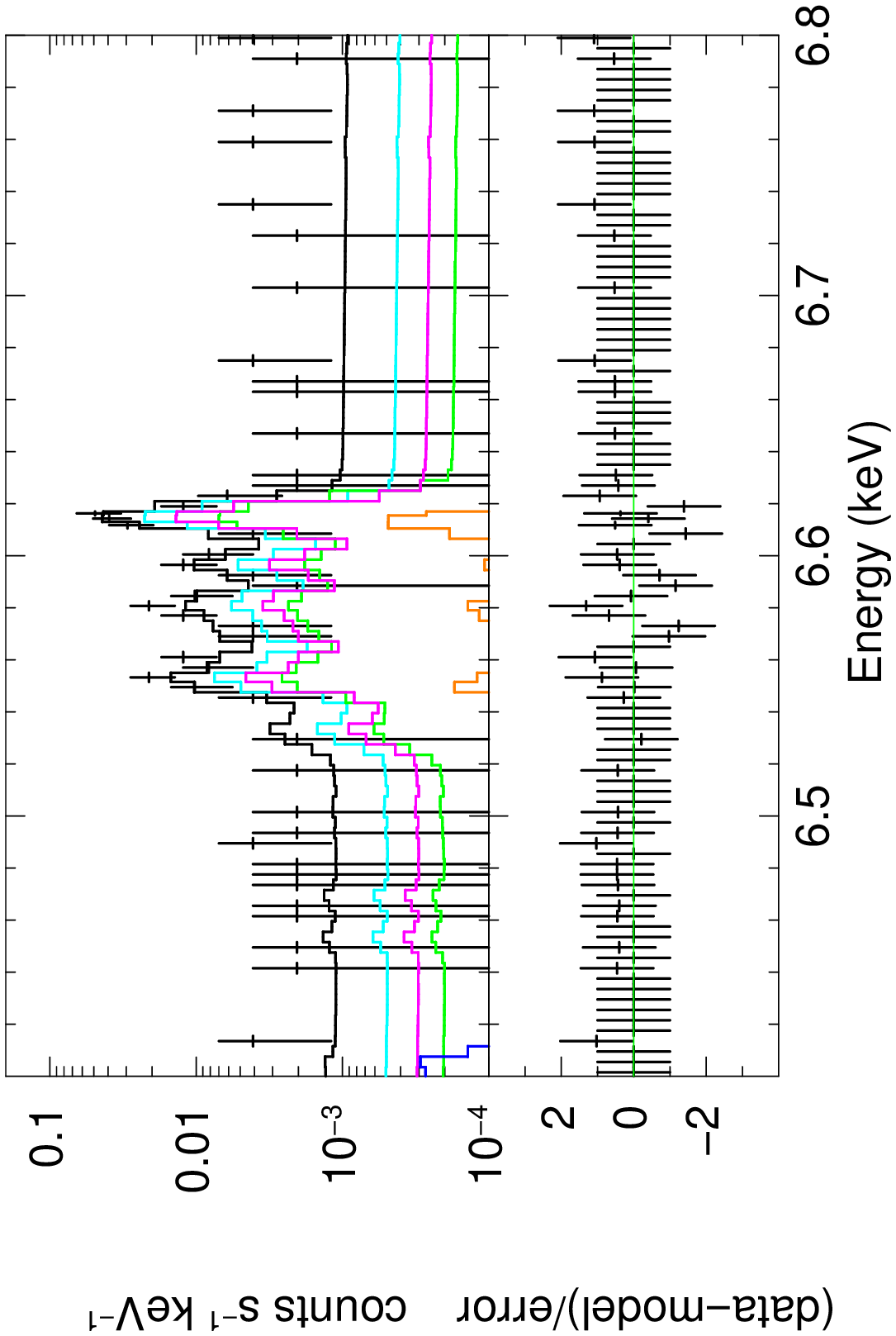}}
\rotatebox{-90}{\includegraphics[width=5.5cm]{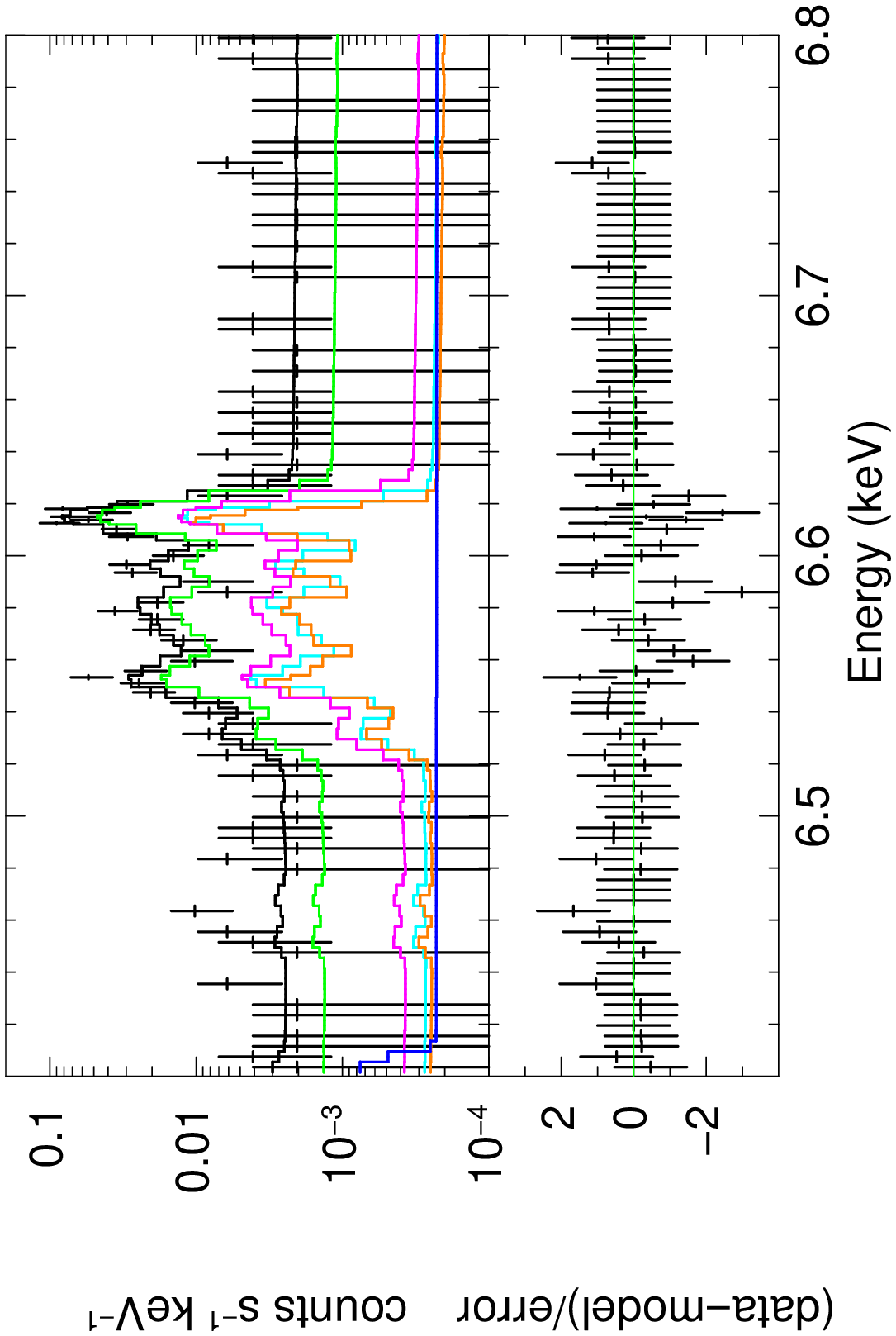}}
\rotatebox{-90}{\includegraphics[width=5.5cm]{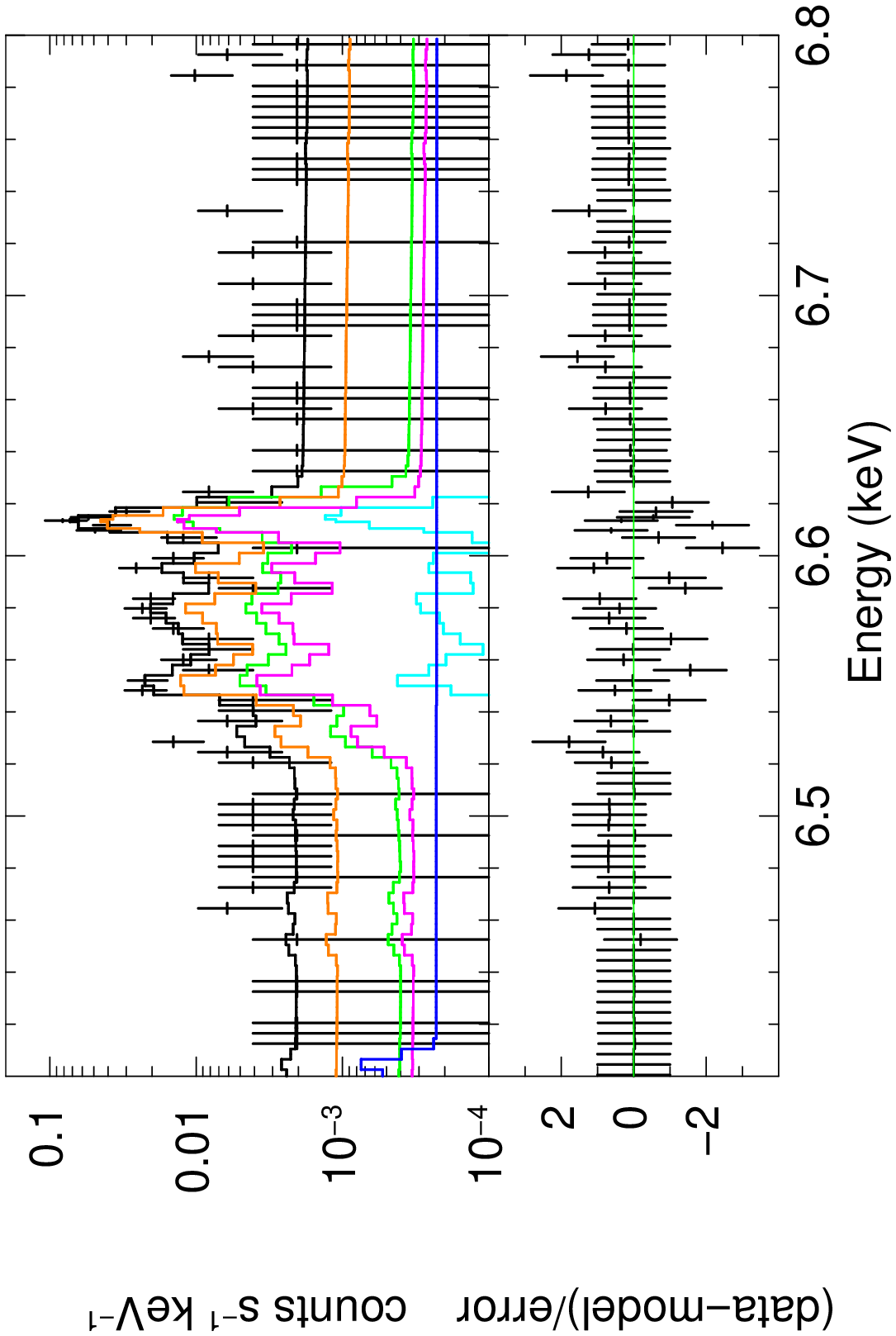}}
\end{center}
\caption{Spectra of each region, shown in order from top to bottom: NW, Center, and SE. Each color corresponds to a different spatial component: cyan represents the NW region, light green the Center region, orange the SE region, and pink the contribution from outside the Resolve pixels. The blue component indicates the NXB. \firstRev[minorRev]{The lower panels show the data-minus-model differences normalized by the statistical uncertainties. Bins with zero observed counts cannot be displayed on the logarithmic upper axes but are included in the lower panels.}\\
{Alt text: Three XRISM Resolve iron-line spectra for the NW, Center, and SE regions. Colored model components show photons originating in each sky region, outside the field of view, and the non-X-ray background; the component matching each extraction region dominates its spectrum.}}
\label{fig:ssm_spectra}
\end{figure}

In the spatially resolved spectral analysis discussed in \firstRev[minorRev]{section}~\ref{sec:Three-region maps}, we applied the Spatial Spectral Mixing (SSM) method by dividing the Resolve FoV into three regions (NW, Center, and SE) to account for photon leakage. \firstRev[minorRev]{Figure}~\ref{fig:ssm_spectra} presents the observed spectra and the corresponding best-fit models for each of these three regions.

The spectra demonstrate that the fits are well constrained and reproduce the observational data accurately thanks to the improved photon statistics. Furthermore, the decomposition of spatial components in each panel visually confirms that the dominant contribution to each spectrum arises from the corresponding region itself, validating our SSM approach and the quantitative mixing fractions presented in the main text.

\section{Dynamo and acceleration efficiencies, and cosmic-ray acceleration and cooling timescales as a function of the turbulence scale}\label{sec:A-1}

\begin{figure}[h]
\begin{center}
\includegraphics[width=\linewidth]{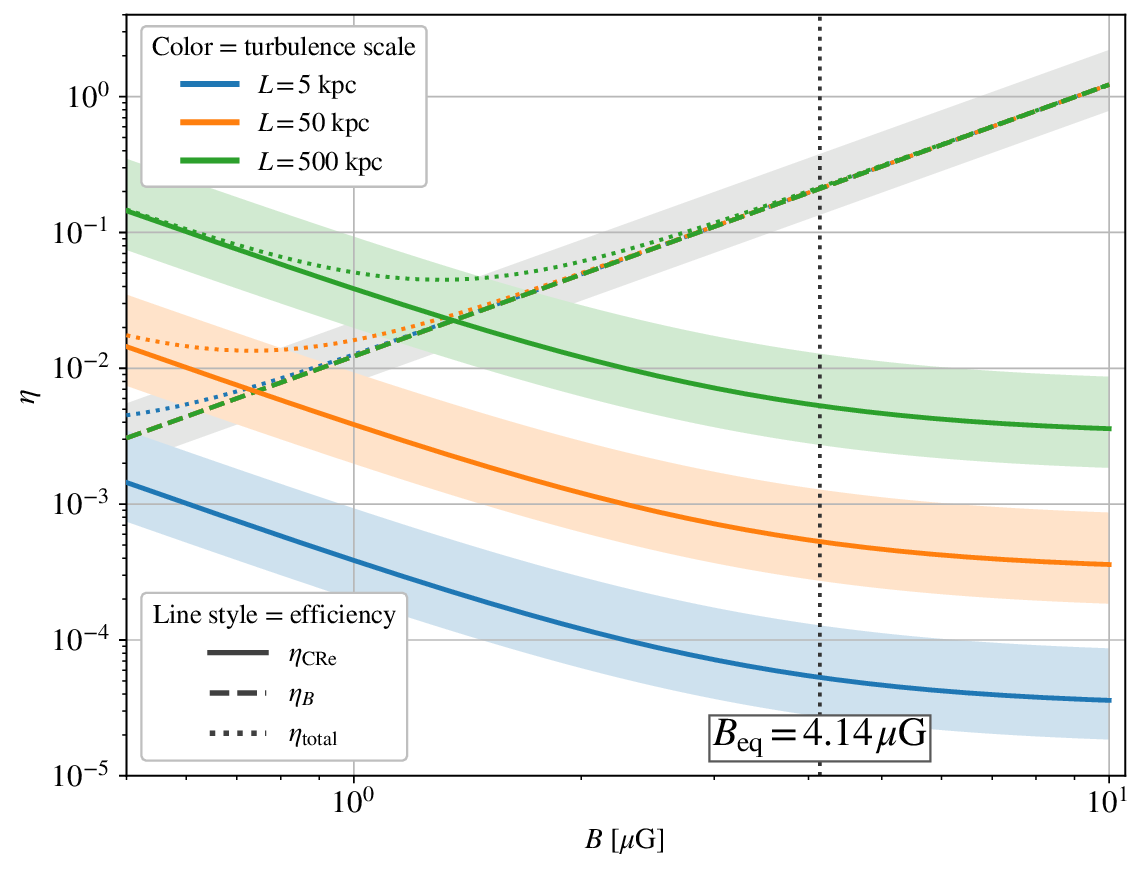}
\end{center}
\caption{
Same quantities and line styles as in \firstRev[minorRev]{figure}~\ref{fig:eta_acc_B}, but for a fixed velocity dispersion of $v_{t,3D}=167.5~{\rm km\,s^{-1}}$ and turbulence scales of $L=5$ (blue), 50 (orange), and 500 kpc (green). \firstRev[minorRev]{Colored bands show the 90\% uncertainty ranges in the CRe efficiency. The gray band shows the corresponding dynamo-efficiency range, which is independent of $L$.} The vertical dotted line marks $B_{\rm eq}=4.14~\mu$G.
\\
{Alt text: Logarithmic acceleration, dynamo, and total efficiency curves versus magnetic field strength for turbulence scales of 5, 50, and 500 kiloparsecs at fixed velocity dispersion. Shaded bands show velocity uncertainties, and a vertical line marks the equipartition field.}
}
\label{fig:eta_acc_B_L}
\end{figure}

\begin{figure}[h]
\begin{center}
\includegraphics[width=\linewidth]{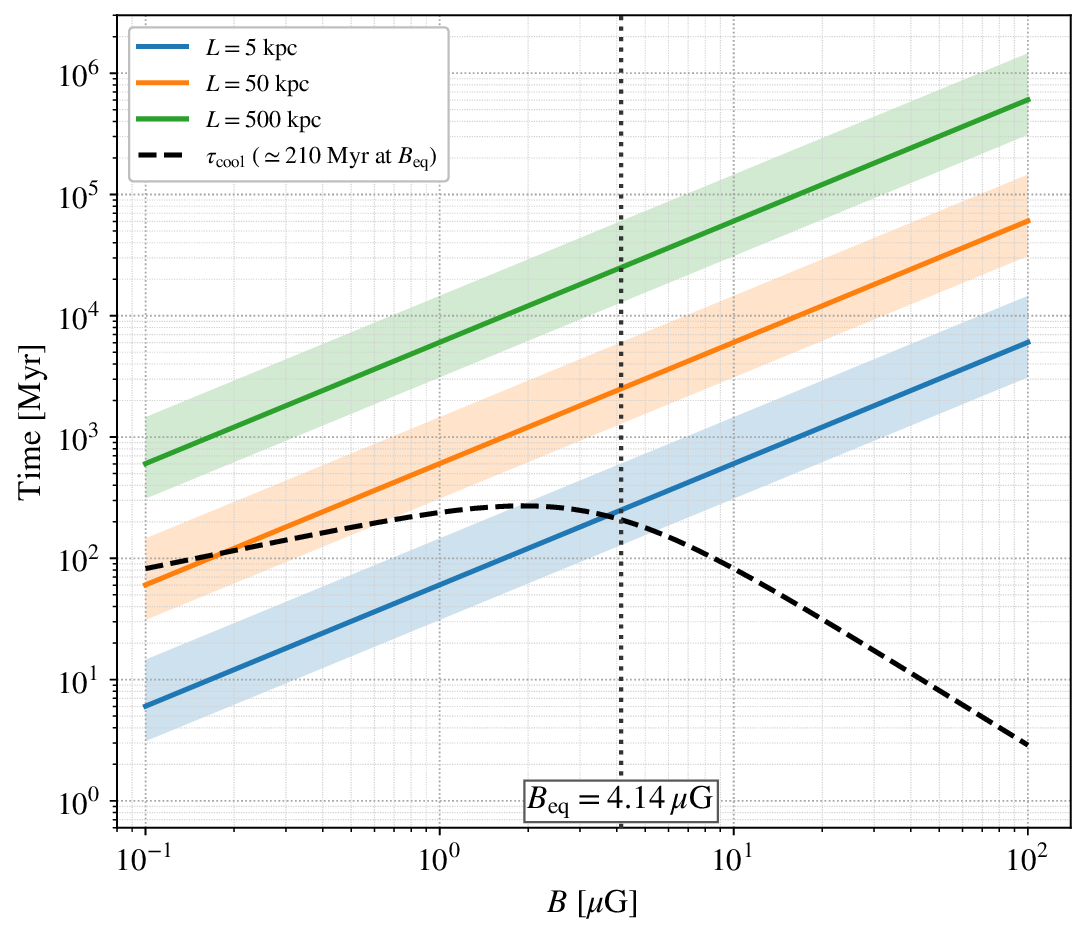}
\end{center}
\caption{
Same quantities and line styles as in \firstRev[minorRev]{figure}~\ref{fig:tau_acc}, but for a fixed velocity dispersion of $v_{t,3D}=167.5~{\rm km\,s^{-1}}$ and turbulence scales of $L=5$ (blue), 50 (orange), and 500 kpc (green). \firstRev[minorRev]{The shaded bands show the propagated 90\% velocity-dispersion uncertainties.} \firstRev{The cooling curve uses the same scaling and adopted normalization as in figure~\ref{fig:tau_acc}.} The vertical dotted line marks $B_{\rm eq}=4.14~\mu$G.
\\
{Alt text: Acceleration and cooling timescales versus magnetic field strength for turbulence scales of 5, 50, and 500 kiloparsecs at fixed velocity dispersion. Colored curves and shaded bands show acceleration times and uncertainties, a black dashed curve shows cooling, and a vertical line marks the equipartition field.}
}
\label{fig:tau_acc_L}
\end{figure}

Here, we investigate the distributions of the dynamo and acceleration efficiencies, as well as the acceleration and cooling timescales, as a function of the turbulence scale. \firstRev[minorRev]{Figures} \ref{fig:eta_acc_B_L} and \ref{fig:tau_acc_L} are analogous to \firstRev[minorRev]{figures} \ref{fig:eta_acc_B} and \ref{fig:tau_acc}, respectively, but with the \firstRev[minorRev]{adopted} turbulence scale $L$ varied. Because $v_t$ enters the relevant relations quadratically and cubically, respectively, even a factor-level change can significantly alter the distributions. In contrast, $L$ enters only linearly, implying that order-of-magnitude variations are necessary to produce comparable effects.

\end{document}